\documentclass[%
 aip,
 jcp,
 amsmath,amssymb,
 reprint,%
]{revtex4-1}
\usepackage{graphicx}
\usepackage{epstopdf}
\usepackage{dcolumn}
\usepackage{bm}

\usepackage{rotating}
\usepackage[utf8]{inputenc}
\usepackage[T1]{fontenc}
\usepackage{amsmath,amssymb}
\usepackage{mathptmx}
\usepackage{etoolbox}
\usepackage{soul}
\usepackage{parskip} 
\usepackage[usenames,dvipsnames]{xcolor}
\usepackage{bigints}
\usepackage{verbatim}
\usepackage{mathrsfs}
\AppendGraphicsExtensions{.tga}

\newcommand{\Dbulk}{D_b}

\makeatletter
\def\@email#1#2{%
 \endgroup
 \patchcmd{\titleblock@produce}
  {\frontmatter@RRAPformat}
  {\frontmatter@RRAPformat{\produce@RRAP{*#1\href{mailto:#2}{#2}}}\frontmatter@RRAPformat}
  {}{}
}%

\makeatother
\begin{document}

\preprint{AIP/123-QED}

\title{Solvent transport near a fluctuating membrane}
\title{Molecular transport enhancement near a fluctuating membrane}
\title{Molecular motion is enhanced near a soft fluctuating membrane}
\title{Molecular motion near a soft fluctuating membrane}
\title{Enhanced molecular diffusion near a soft fluctuating membrane}

\author{Ali Mohammadi}
\affiliation{Department of Mechanical Engineering, Clemson University, Clemson, SC 29634, USA}
\author{Zhen Li}
\affiliation{Department of Mechanical Engineering, Clemson University, Clemson, SC 29634, USA}

\author{Sophie Marbach}
\email{sophie.marbach@cnrs.fr}
\affiliation{CNRS, Sorbonne Université, Physicochimie des Electrolytes et Nanosystèmes Interfaciaux, F-75005 Paris, France}

\author{Micheline Abbas}
\email{micheline.abbas@toulouse-inp.fr}
\affiliation{Laboratoire de Génie Chimique, Université de Toulouse, CNRS, INPT, UPS, Toulouse, France}

\date{\today}

\begin{abstract}

Particles diffusing near interfaces face anisotropic resistance to motion due to hydrodynamic interactions. While this has been extensively studied near \textit{hard} interfaces since the works of Lorentz and Brenner, our understanding of diffusion near \textit{soft, thermally fluctuating} interfaces remains limited. Previous studies have predominantly focused on particles much larger than the molecular scale at which thermal fluctuations become important. In this work, we numerically investigate the dynamics of individual solvent molecules near a thermally fluctuating lipid membrane, a canonical soft interface in biology. We observe that diffusive motion of solvent molecules near the fluctuating membrane is slightly enhanced compared to a flat rigid interface and significantly more so than near an undulated rigid interface. This enhancement in diffusive motion of solvent molecules arises from spontaneous momentum exchanges between the moving membrane and adjacent molecules, promoting mixing. Notably, this dispersion effect overcomes geometric trapping that slows diffusion near the rigid undulated interface. Our analysis reveals that the momentum transfer near the fluctuating membrane is so efficient that it resembles an effective slip boundary condition over a length scale equal to the fluctuation height. These molecular-scale mechanisms differ from those of larger particles, where hydrodynamic memory and elasticity effects can be at play as they relax over timescales comparable to significant diffusive motion. Our findings advance understanding of enhanced diffusive motion and promoted mixing near soft fluctuating membranes involved in diverse biological processes and soft-matter technologies containing natural and model cell membranes.
\end{abstract}

\maketitle

\section{Introduction} 
The diffusion of dispersed species near interfaces plays a critical role in many interfacial phenomena such as adsorption and capture. In biological systems, for instance, the interplay between diffusion and intermittent adsorption strongly influences protein binding rates to cells and other targets~\cite{nguyen2024competition,lomholt2007subdiffusion}. Similar competition between diffusive transport and physicochemical interactions is also relevant in technological applications, including surface-based detection sensors~\cite{katelhon2016near} and film formation processes in coatings~\cite{kaiser2002review}.
In an unbounded viscous fluid, the diffusion of dilute particles is characterized by their self-diffusion coefficient \(\Dbulk\). Einstein's relation connects $\Dbulk$ to a friction coefficient $\gamma$, as $\Dbulk = k_B T/\gamma$ with $k_BT$ representing thermal energy. For a spherical particle of radius \(a\) substantially larger than fluid molecules, freely moving in an unbounded fluid at low Reynolds numbers, the friction coefficient in a fluid with dynamic viscosity \(\eta\) is given by Stokes' law~\cite{stokes1851effect}, $\gamma = 6 \pi \eta a$, giving the Stokes-Einstein relation $\Dbulk = k_B T /6 \pi \eta a$.

In the presence of an interface, hydrodynamic interactions between the particle and the interface increase the friction coefficient $\gamma$, and thus decrease diffusion. Several analytical expressions have been obtained near \textit{hard, flat} interfaces.
Lorentz obtained pioneering expressions for the diffusion of a spherical particle moving near a rigid wall~\cite{lorentz1907entstehung}, using a method of images with a first-order reflection. 
When the particle’s radius \( a \) is much smaller than its instantaneous distance \( z \) from the wall, \( a/z \ll 1 \), the local diffusion coefficients may be approximated by 
\begin{equation}
    \begin{split}
        \frac{D_\perp(z)}{\Dbulk} \approx 1 + \frac{9}{8}  \frac{a}{z}  \hspace{2mm} \text{and} \hspace{2mm} 
         \frac{D_\parallel(z)}{\Dbulk} \approx 1 + \frac{9}{16} \frac{a}{z} 
    \end{split}
    \label{eq:lubrication}
\end{equation}
for the perpendicular and parallel directions with respect to the wall. Wakiya~\cite{wakiya1960} and Fax\'{e}n~\cite{faxen1921einwirkung,oseen1927neuere} improved the Lorentz results by incorporating the effect of the second-order reflection~\cite{bian2016111}. For the perpendicular direction, Brenner~\cite{brenner1961slow} derived the exact solution to the creeping equations for a spherical particle moving near a rigid wall, unrestricted by the ratio \(a/z\). While Brenner's solution is expressed as an infinite series, approximations have been obtained through regression methods~\cite{honig1971effect,bevan2000hindered}.
For the parallel case, an exact solution valid for the entire range of \( z \) does not exist. Goldman~\cite{goldman1967slow} and O'Neil~\cite{o1967slow} derived asymptotic solutions in the limit as \( z \rightarrow 0 \). Perkins~\cite{perkins1992hydrodynamic} matched the series solutions to the asymptotic results (\( z \rightarrow 0 \)) to find the friction coefficient for the entire range of \(z\). Both Perkins' and Brenners' solutions were found to agree remarkably with experimental data~\cite{carbajal2007asymmetry}.

However, in many contexts, especially in biology, interfaces are far more complex, being neither hard, flat, nor static. Lipid membranes, which serve as critical molecular barriers between cells, represent a canonical example of complex interfaces due to their intrinsic softness, lipid fluidity, and thermal fluctuations~\cite{quemeneur2014shape,girard2005passive,brandt2011interpretation}. Nevertheless, our understanding of how particle motion is influenced by proximity to such soft, thermally fluctuating interfaces is still in its infancy. In particular, a clear unknown is the shape of the diffusion coefficient $D_{\perp/\parallel}(z)$ as a function of the distance to the fluctuating interface $z$.

Recent investigations have revealed novel transport mechanisms for particles significantly larger than the molecular scales at which thermal fluctuations occur. In confined systems such as pores, theoretical and experimental studies have demonstrated that diffusion is strongly influenced by the characteristics of the fluctuating interface, leading to enhanced diffusion relative to bulk conditions, particularly for rapid or large-wavelength fluctuations of the pore walls~\cite{marbach2018transport,wang2023interactions,marbach2019active,sarfati2021enhanced,chakrabarti2020}. This enhanced diffusion arises from longitudinally fluctuating fluid flows, driven by the fluctuating channel walls~\cite{marbach2018transport}, a phenomenon reminiscent of Taylor-Aris dispersion~\cite{taylor1953,aris1956dispersion,marbach2019active}. Yet when a particle is close to only one fluctuating interface, diffusion can be slowed down and the particle's motion can become subdiffusive over timescales smaller than the interface relaxation timescale~\cite{sheikh2023brownian}. Those observations suggest particle size plays a critical role in confined fluctuating systems. In addition, to fully understand the transport of smaller species, such as ions, molecules, or nanoparticles—particularly relevant in biomedical contexts—it is crucial to examine how molecular-scale transport is affected.

At molecular scales, since thermal fluctuations become significant, it is natural to expect stronger and qualitatively different transport phenomena~\cite{kavokine2021fluids}. Theories for nanoscale hydrodynamics near interfaces are generally complex and require significant number of closures~\cite{camargo2018nanoscale}, so molecular transport processes have mostly been studied numerically. Early molecular dynamics simulations revealed that gas diffusion in microporous materials is markedly enhanced when thermal vibrations of the solid matrix are taken into account~\cite{leroy2004self,haldoupis2012quantifying}; the resulting increase can span several orders of magnitude, far exceeding the perturbative effects observed at larger scales~\cite{fares2024observation,sarfati2021enhanced}. Similar enhancements in solvent diffusion within solid-state pores have been attributed to phonon-fluid couplings~\cite{ma2015water,ma2016fast,cao2019water,noh2021phonon} or extreme geometrical confinement~\cite{yoshida2018dripplons}. Conversely, hydrophilic interactions between water molecules and lipid interfaces can create local traps that suppress lateral transport~\cite{sedlmeier2011water,von2013anomalous}. These studies rely on molecular dynamics simulations that explicitly incorporate the detailed molecular interactions specific to each system. This raises a broader question: what are the general principles governing molecular transport near unconfined, soft, fluctuating membranes? How do these mechanisms connect to the well-established lubrication phenomena observed at larger scales, as \textit{e.g.} described by Eq.~\eqref{eq:lubrication}?

In this work, we investigate the dynamics of solvent molecules near a model lipid bilayer membrane that exhibits long-range thermal fluctuations. To uncover general transport behavior, we employ mesoscale simulations based on dissipative particle dynamics (DPD), which is rooted in atomistic dynamics and can be derived directly from molecular dynamics using rigorous Mori-Zwanzig projection~\cite{li2014construction}. This approach enables us to simultaneously resolve molecular motion and capture long-wavelength membrane deformations~\cite{gubbiotti2022electroosmosis}. In our simulations, the space above the interface is divided into discrete layers, and statistical analyses are performed separately for particles based on the layer in which they were initially located. This allows us to measure transport properties as a function of the distance from the membrane. We mainly observe an enhancement of local diffusion in both directions near the fluctuating membrane, compared to  a flat rigid wall or a wavy rigid membrane interface. Long time molecular diffusion parallel to the fluctuating membrane is also enhanced. We show that this fluidification effect is mostly due to the dynamic ``wiggling'' of the fluctuating membrane, by investigating several potential mechanisms. Remarkably, dynamic fluctuations transfer enough momentum to the solvent to facilitate mixing and overcome the geometrical ``trapping'' effect induced by the presence of wells in the membrane. Although such dynamic wiggling was shown to enhance long time, global diffusion properties only theoretically~\cite{marbach2018transport}, here we demonstrate this effect on \textit{local} diffusion and show that it is significantly larger at the molecular scale than predicted by theory.

\section{Models and System Setup}
\subsection{A fluctuating membrane model}



\begin{figure}[h!]
    \centering
    \includegraphics[scale=0.21]{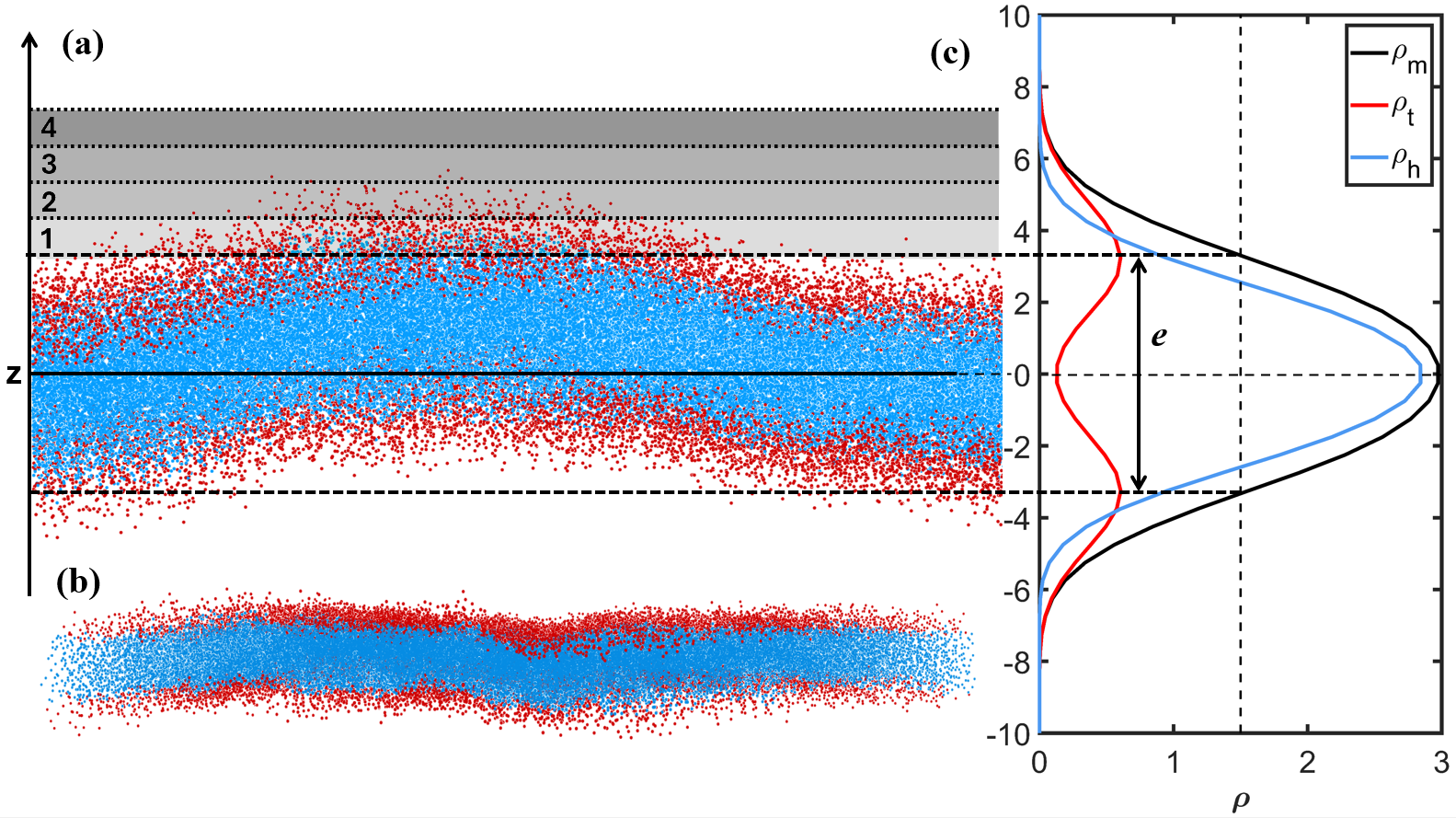}
    \caption{(a) 2D zoomed-in cross-sectional snapshot of the fluctuating membrane, including the slabs where statistics have been performed. Points represent DPD beads, with hydrophilic heads and hydrophobic tails in red and blue, respectively. (b) Zoomed-out configuration. (c) Time-averaged density profiles with height $z$ ($\times \, \ell_0$) corresponding to the membrane (black), heads (red), and tails (blue). Dashed lines (extended from (a) to (c)) show the average thickness of the membrane \(e\), \textit{i.e.} points where the membrane density is equal to $1.5 \ \ell_0^{-3}$. All spatial dimensions are in  $\ell_0$ unit and $\rho$ is given in $\ell_0^{-3}$}   \label{fig:membrane_density}
\end{figure}

Our model membrane consists of a lipid bilayer assembly, freely fluctuating in a solvent representing water. 
We use a model for phosphatidylethanolamine lipids, that are frequently found in biological membranes. This phospholipid has one hydrophilic head (red dots in Fig.~\ref{fig:membrane_density}) and two hydrophobic tails (blue dots).
This particular choice is not expected to affect the generality of our results. 
Indeed, Kranenburg et al.~\cite{kranenburg2004comparison} found that a bilayer can always be obtained for various DPD parameters as long as the phospholipid has well-defined hydrophilic and hydrophobic parts. Further details are given in Appendix~\ref{app:coarse_grain}.


The fluctuating hydrodynamics simulations are based on the DPD method\cite{groot1997dissipative,espanol1995statistical}, a particle-based Lagrangian method in which each particle represents a coarse-grained cluster of fluid molecules. Each DPD particle indexed by $i$ is characterized by its position $\mathbf{r}_i$ and momentum $m\mathbf{v}_i$, and its time evolution is governed by Newton's second law of motion. 
Forces on particles are determined through pairwise interactions with all other particles within a cutoff distance $\ell_0$. These forces include conservative forces, in particular bond interactions to link the particles in the lipid chain. Dissipative and random forces jointly act as a thermostat, satisfying the fluctuation-dissipation theory. Repulsive force parameters are set to match the compressibility of the solvent to that of liquid water~\cite{groot1997dissipative}.
For particles of different types, force parameters are determined to reproduce the mutual solubility of different species by matching the Flory–Huggins parameters~\citep{groot1997dissipative}. More details are given in Appendix~\ref{app:dpd}.

As system parameters, we define $\ell_0$ the unit length, $k_BT$ the unit of energy and $\tau_0 =\ell_0 \sqrt{m/k_BT}$ the unit time, where $m$ is the mass of a bead (all the beads having the same mass). For our model system of a lipid bilayer immersed in water, $\ell_0=0.64 \ nm$ and $\tau_0=3\ ps$. 
Every DPD bead roughly occupies the same volume as three water molecules. While there is no inherent radius $a$ for a bead, we define twice the molecule radius $2a$ as the average distance between beads, obtained from the maximum of the solvent radial distribution function, giving $a = 0.41 \ell_0$. 
The simulation box size is $(L_x\times L_y\times L_z)=(90.7\times 90.7\times 46)\ell_0^3$. 
With a number density $\rho=3 \ \ell_0^{-3}$, the total number of DPD beads in this system is approximately \( 1,135,000 \), with approximately $10\%$ of membrane beads.
The coordinate perpendicular to the membrane is $z$ (or $\perp$), whereas $x$ and $y$ (or $\parallel$) indicate interchangeably the directions parallel to the membrane. The membrane size is set to match the simulation box size, to keep the time-averaged surface tension sufficiently low. All numerical parameters are recapitulated in Appendix~\ref{app:dpd_param}. 

We perform simulations using the open-source code LAMMPS \citep{LAMMPS}. A modified velocity-Verlet algorithm \citep{groot1997dissipative} is used to integrate the DPD governing equations with time step $\Delta t = 0.01\ \tau_0$. 
Simulations including a membrane are carried out using a pre-assembled bilayer, after an initial thermal relaxation of the membrane for \( 10^5 \) time steps. Because the coarse-grained DPD simulation still preserves the molecular structure of the fluid, from then on we refer to solvent beads as solvent molecules.

\vspace{-4.5mm}
\subsection{Structural membrane properties}

\begin{figure}[th]
\centering
\includegraphics[scale=0.6]{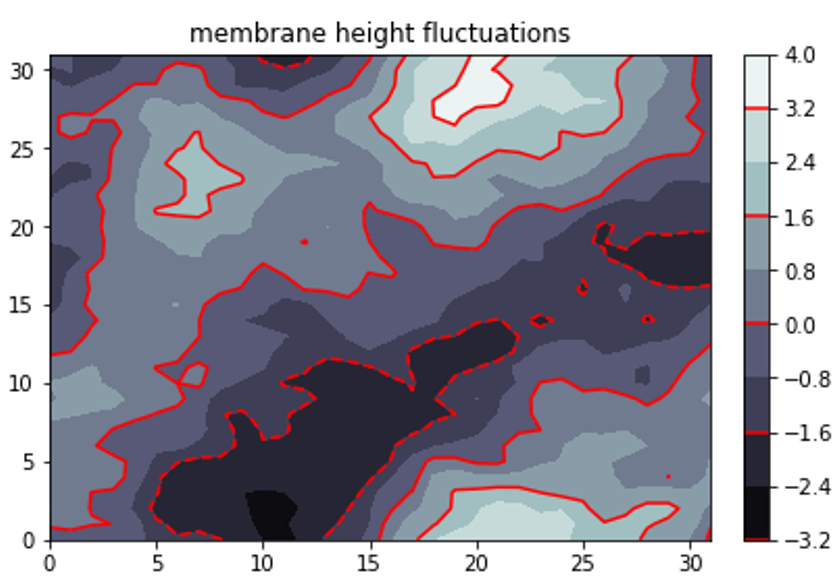}
\caption {Snapshot showing membrane height fluctuations, projected on a lateral $(32\times 32)$ grid at time $t=1000\ \tau_0$. The color bar indicates the local height with respect to the membrane midplane in $\ell_0$ unit. }
\label{fig:mb_snapshot}
\end{figure}

Fig.~\ref{fig:membrane_density} shows a 2D cross-sectional snapshot of the membrane and its time-averaged density profile. The thickness $e$ of the membrane is determined by locating the \(z\)-coordinates for which the time-averaged number density is half the maximum value; here, \( e = 6.70 \, \ell_0 \). At each point in time, we project lipid tail beads on a $32\times 32$ 2D grid parallel to the membrane, and calculate their average instantaneous height $h(\bm{x},t)$ per grid square, relative to the membrane midplane.
The root-mean-square height fluctuations of the membrane are $\delta h = \sqrt{\langle(h(\bm{x},t)- \langle h(\bm{x},t) \rangle)^2 \rangle}\approx 1.3 \,\ell_0$, where the brackets $\langle \cdot \rangle$ indicate an average over time and grid squares. Fig.~\ref{fig:mb_snapshot} provides a snapshot of membrane height fluctuations at equilibrium.

To characterize the membrane "static" properties, we followed a method described in our previous work \cite{sheikh2023brownian}, using simulations carried out for $10^4\tau_0$. Investigating the static structure factor of the height fluctuations, we measured a surface tension $\sigma=0.006\ k_BT/\ell_0^2$ and a bending rigidity $\kappa\approx 10\  k_BT$. We also measured a stretching modulus $K_A\approx9k_BT/\ell_0^2$ based on the linear increase of the membrane surface tension with the area per lipid. Here, $\kappa$ and $K_A$ are smaller than in our previous work~\cite{sheikh2023brownian} because the solubility mismatch between solvent and hydrophobic beads is smaller here ($\Delta a_{ij}=26$ compared to $\Delta a_{ij}=55$).

\subsection{Statistical analysis per slab}

In this work, we focus on solvent dynamics as a function of the distance to the interface. One common challenge for such data analysis is that the solvent residence time at a certain distance from the membrane is short, reducing statistical accuracy. To circumvent this it is common to perform statistics on particle trajectories that start in a given layer (conditional probability)
~\cite{vilquin2021time}. We thus divide the simulation domain into slabs parallel to the interface, as sketched in Fig.~\ref{fig:membrane_density}. The slab thickness is equal to $\ell_0$, which we recall corresponds to the cut-off distance for the DPD forces. 
A slab's thickness is thus close to the intermolecular distance $2a \simeq 0.82 \ell_0$. Thus, a solvent molecule meets several neighbors during its stay in a slab.
The bottom edge of the first slab corresponds to the coordinate $z$ where the number density of the solvent is equal to half that in the bulk, $1.5\ell_0^{-3}$ (dashed lines in Fig.~\ref{fig:membrane_density}c).
In each simulation, the statistical quantities computed in a given slab are performed on the molecules that were located in that slab at start of the simulation -- but after the thermalization process. 

\subsection{Comparison with other interfaces}

To disentangle the effects of membrane shape and dynamics, we conducted simulations of solvent dynamics near three interfaces with identical simulation domains: (a) the fluctuating membrane, (b) a rigid undulated membrane, and (c) a flat rigid wall. To prepare (b), the fluctuating membrane beads are frozen after an initial relaxation (\( 10^5 \) time steps) by excluding them from time integration and setting their velocity to zero.
The flat rigid wall in case (c) was generated by randomly placing wall beads with a similar number density of \( \rho = 3 \ \ell_0^{-3} \) within a ``wall'' volume spanning the simulation domain horizontally and with a thickness $e = 6.7 \ell_0$ similar to the thickness of the fluctuating membrane. After a relaxation process of \( 5000 \) time steps, the wall beads were excluded from time integration, and their velocities were set to zero. The interaction parameters between the solvent molecules and the flat wall beads were set equal to that between the solvent molecules and the hydrophilic lipid beads. 
As a consequence, the solvent slightly penetrates the flat wall after the thermalization process, akin to the penetration that occurs on the surface of the membrane.

\begin{figure}[th]
\centering
\includegraphics[scale=0.65]{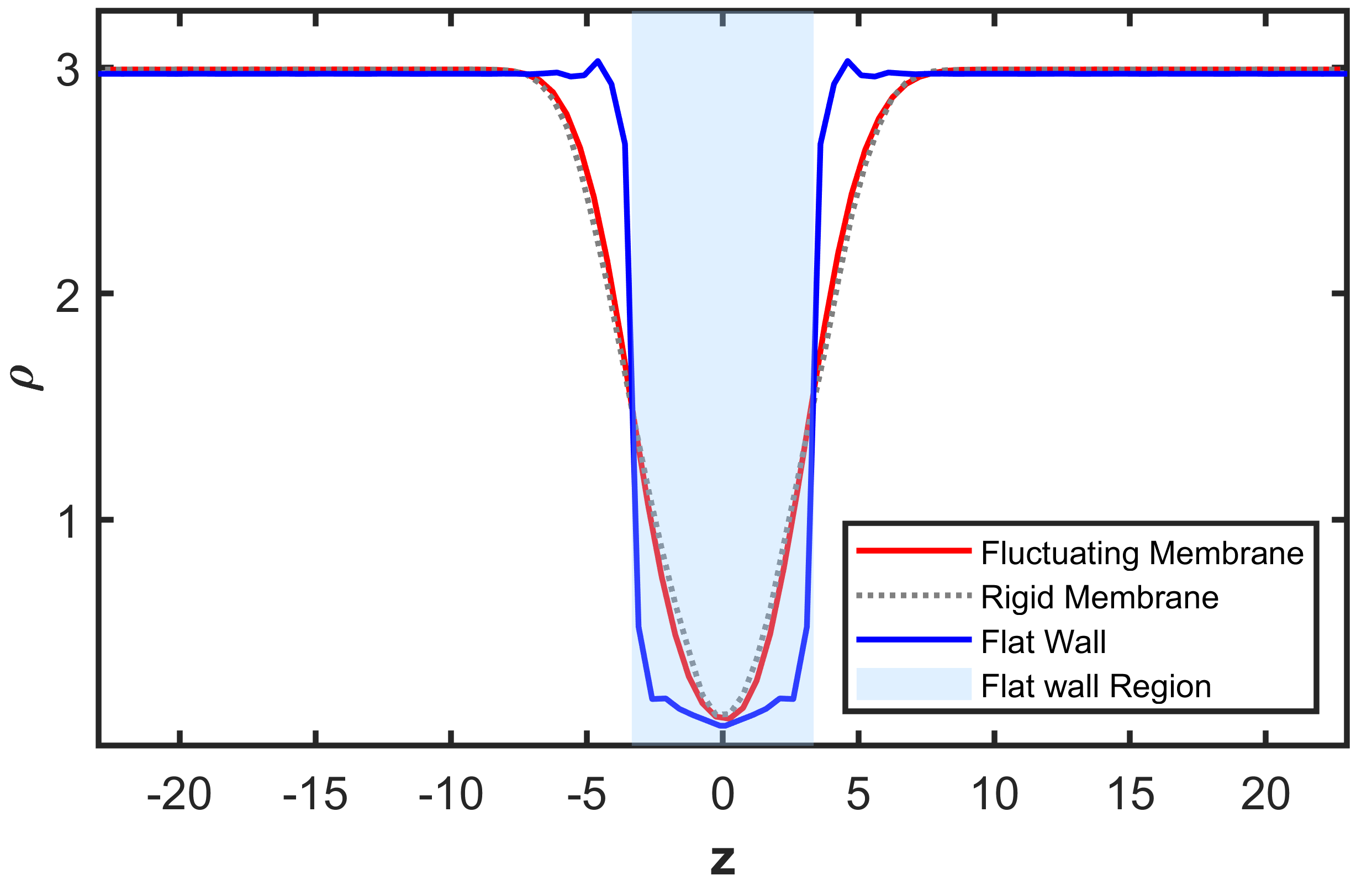}
\caption{Time-averaged density profile of solvent molecules in the presence of a fluctuating membrane (solid red), flat rigid wall (dashed blue) or a rigid membrane (dotted gray). The coordinate $z$ is indicated in $\ell_0$ unit. Gray and red lines have significant overlap. The shaded region indicates the thickness of the flat wall and the average position of the membrane surfaces, where the density of solvent molecules is half that in the bulk.}
\label{fig:w_density_profiles}
\end{figure}

Fig.~\ref{fig:w_density_profiles} shows the solvent density profiles, averaged along the parallel directions and over time, for the 3 different cases. Profile points are obtained along $z$ within slabs of thickness $0.5\ \ell_0$. The solvent density profile varies sharply near the flat interface, indicating a clear separation between the solid and fluid phases.  In comparison, the solvent density profiles vary smoothly near the membrane surfaces since the solvent can occupy open volume regions within wells of the membrane, while it is evacuated from the bumpy regions. The profiles are almost identical for the fluctuating (in red) and rigid (in dashed gray) membranes, which indicates that dynamic events associated with the membrane fluctuations do not impact the solvent density on average. 

We now turn to examine the dynamics of solvent molecules.

\section{Brownian motion of the solvent molecules is enhanced near a fluctuating membrane.} 
\label{sec:observations}

\subsection{Mean-squared displacement of solvent molecules: from ballistic to diffusive} 

\begin{figure}[h!]
\centering
\includegraphics[scale=0.22]{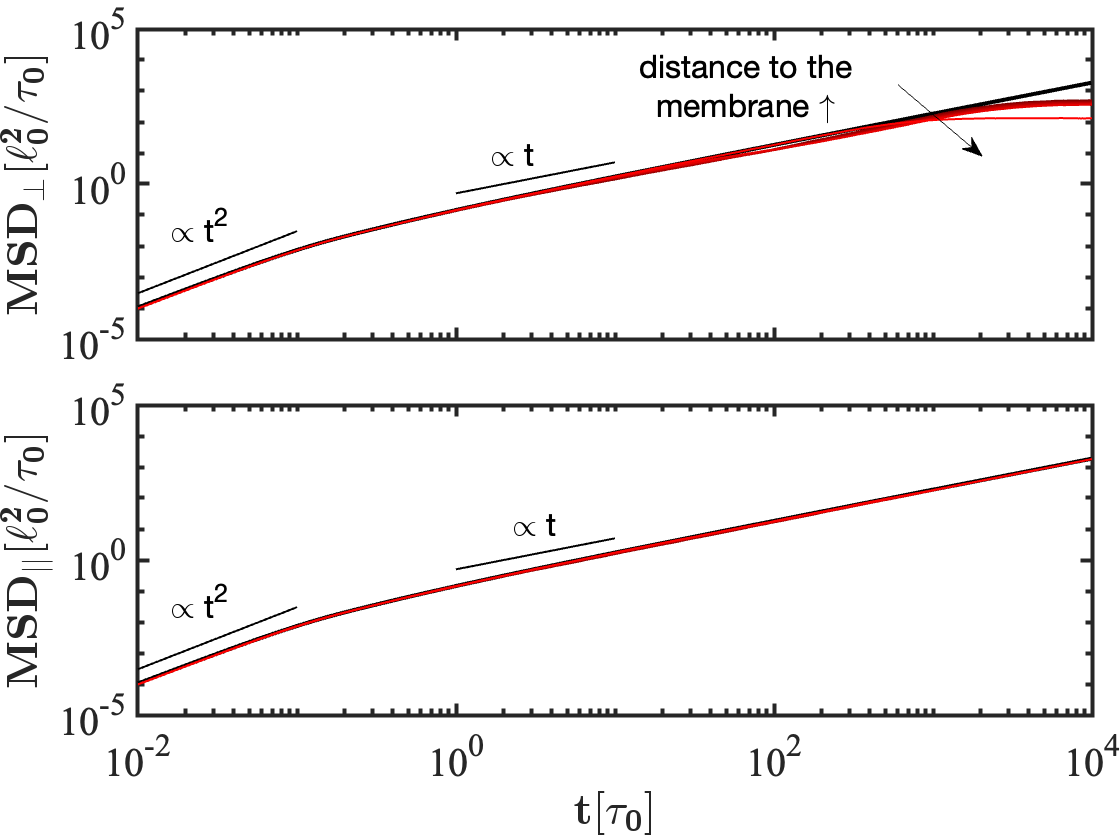}
\put(-210,170){(a)}
\put(-210,85){(b)}
\caption{Mean square displacement (MSD) of solvent molecules near the membrane in the perpendicular (a) and parallel (b) directions. Red lines (almost superimposed), from darkest to lightest, represent slabs at distances of 0.5, 1.5, 2.5, 3.5, 4.5 and $19.5\,\ell_0$ from the fluctuating membrane surface. Black lines indicate the bulk case.} 
\label{fig:MSD}
\end{figure}

To characterize Brownian motion of the solvent molecules near the different interfaces, we first consider their mean-square displacement (MSD) in each direction as
\begin{equation}
\text{MSD}_{\perp,\parallel}(t) = \langle |\mathbf{r}_{\perp,\parallel}(t) - \mathbf{r}_{\perp,\parallel}(0)|^2 \rangle,
\end{equation}  
where $\langle \cdot \rangle$ denotes an average on molecules and on moving time windows.  
Fig.~\ref{fig:MSD} displays the MSD of the molecules in the perpendicular and parallel directions near the fluctuating membranes. The different shades of red correspond to the different slabs, so to molecules which started their motion at different distances from the fluctuating membrane. Similar profiles are obtained near the flat and rigid membranes (not shown here). 
As reference we also show the MSD for a bulk simulation in black. 

Fig.~\ref{fig:MSD} shows that the MSDs follow 2 different power-laws in time, which are apparent for all slabs and for the bulk case, in both direction. Motion is ballistic and then around $t \simeq \tau_{\rm diff} \simeq 1 \ \tau_0$, transitions to a diffusive regime.  
At long times, $\text{MSD}_\perp(t)$ tends to a plateau, a signature of confinement between the interface and its periodic image. 
From the time scale $\tau_{\rm conf} \simeq 10^3 - 10^4\ \tau_0$ at which the plateau emerges, we conclude that the investigation of the diffusive dynamics in the perpendicular direction is appropriate for $t < \tau_{\rm conf}$. 
At long times, we can quantify the self-diffusion coefficient of the solvent in the parallel direction, as 
\begin{equation}
    D_{\parallel}^{\infty} = \lim_{t\rightarrow \infty} \frac{\text{MSD}_{\parallel}(t)}{2t}.
\end{equation}

For the bulk case, motion is isotropic and we find $\Dbulk \equiv D_{\parallel}^{\infty} = 0.092 \ \ell_0^2/\tau_0\). 
The time to diffuse the characteristic distance $2a$ between two molecules is $(2a)^2/\Dbulk \simeq 7 \ \tau_0$, which is comparable to the time required to transition to the diffusive regime $\tau_{\rm diff}$. The time to diffuse the vertical distance between periodic membranes is $(L_z - e)^2/\Dbulk \simeq 10^4\ \tau_0$ which is indeed comparable to $\tau_{\rm conf}$.

Already at this level of description, the molecular nature of the investigation is apparent.
The Stokes-Einstein relation in the bulk estimates a diffusion coefficient $k_B T / 6 \pi \eta a \simeq 0.046 \ \ell_0^2/\tau_0$, where the dynamic viscosity $\eta = m \rho \nu$ and \(\nu = 0.94\ \ell_0^2/\tau_0\) was measured from non equilibrium simulations under an applied shear stress (see Appendix~\ref{app:bulk_prop}). This estimate is 2 times smaller than the measured $\Dbulk$, a result which was also obtained in DPD simulations by Pan et al.~\cite{pan2008hydrodynamic} or in MD simulations~\cite{alley1983generalized}. 
This discrepancy may not be explained by inertial effects since the Schmidt number \(\textnormal{Sc} = \nu/\Dbulk = 10.18\) is sufficiently high that momentum in the fluid diffuses much faster than the fluid molecules themselves~\cite{balboa2013stokes}. Rather, the discrepancy is attributable to molecular effects, since the Stokes-Einstein relation is only valid in the continuum limit. Modifications of the relation to account for the molecular fluid nature, for instance attributed to wave-number dependencies of the viscosity coefficient or effective slip at the particle's surface could yield a modified Stokes-Einstein estimate closer to the measured value~\cite{alley1983generalized,vergeles1996stokes,bocquet1994brownian,alder1981validity}. 

\begin{table}[h!]
    \centering
    \renewcommand{\arraystretch}{1.2} 
    \setlength{\tabcolsep}{15pt} 
    \begin{tabular}{c c c}
        \hline \hline
        ~ & $D_{\parallel}^{\infty}[\ell_0^2/\tau_0]$ &  $D_{\parallel}^{\infty}/D_b$ \\ 
        \hline
        fluctuating membrane & 0.0897 & 0.975  \\
        flat wall & 0.0875 & 0.951 \\
        rigid membrane & 0.0859 & 0.934 \\
        \hline \hline
    \end{tabular}
\vspace{-3mm}
    \caption{Diffusion coefficient of the solvent molecules $D_{\parallel}^{\infty}$, in the presence of an interface.}
    \label{table:Db_confined}
\end{table}

Table~\ref{table:Db_confined} summarizes the value of $D_{\parallel}^{\infty}$ near different interfaces, averaged over the six closest slabs to the interface.  Note that at long times, the molecules would have explored the space between the interface and its periodic image. 
Overall, confined molecules are less mobile than in the bulk -- the interface contributes to an increased friction which reduces the diffusion coefficient.  
This reduction (the plateau value) depends on the system considered: the diffusion is less reduced in the fluctuating membrane case than by the rigid membrane. The fluctuating membrane thus appears to ``fluidify'' motion near its interface.

We now turn to a more detailed time analysis.

\subsection{Time-dependent diffusive properties as molecules navigate close to the interface}

To explore the mobility of the molecules near the interface, while excluding the average drift away from the interface (discussed in Sec.~\ref{sec:displacements}), we calculate their local, time-dependent, diffusion, through the expression
\begin{equation}
    D_{\perp,\parallel}(t) = \frac{1}{2}  \frac{\sigma_{\perp,\parallel}^2(t)}{t}
    \label{eq:Dperp}
\end{equation}
where $\sigma_{\perp,\parallel}^2(t)$ is the standard deviation (STD) of the molecules' displacements in each slab, \textit{i.e.} where any short time drift contribution is removed. 

\begin{figure}[h!]
\centering
\includegraphics[scale=0.22]{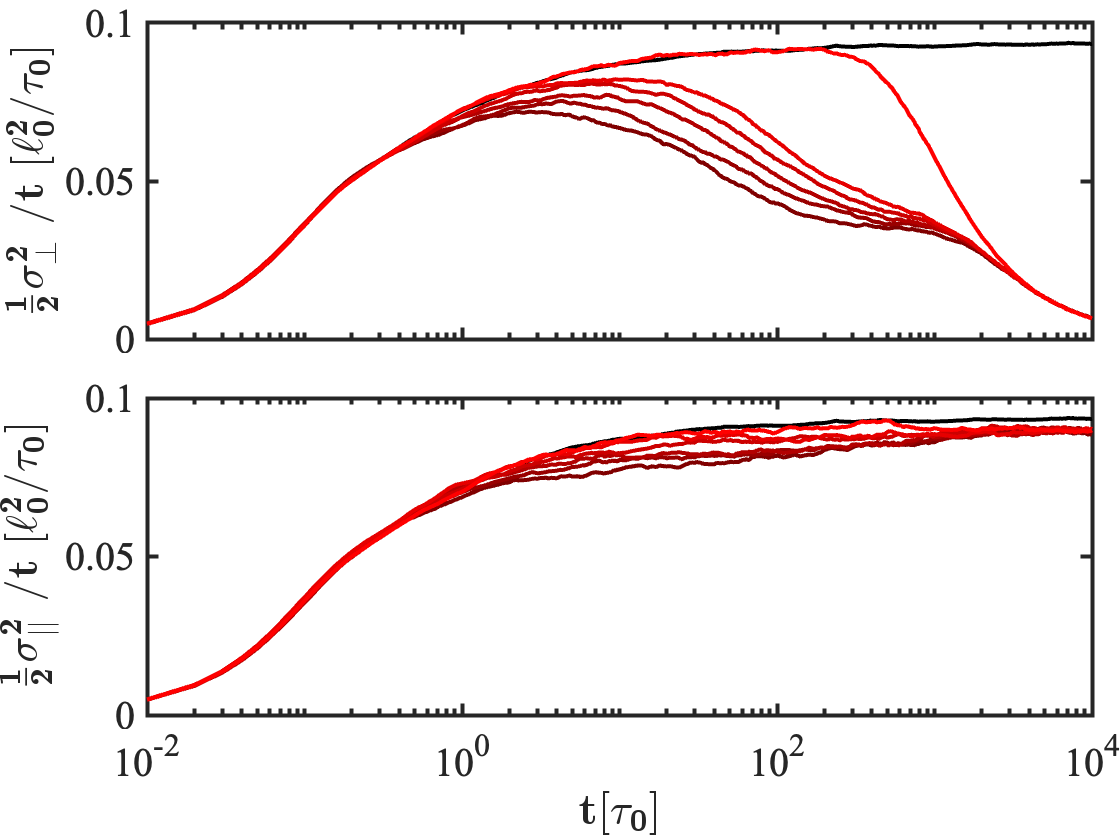}
\put(-200,170){(a)}
\put(-200,85){(b)}
\caption{Time evolution of $\frac{1}{2}  \frac{\sigma_{\perp,\parallel}^2(t)}{t}$ in the (a) perpendicular and (b) parallel directions. Red lines, from darkest to lightest, represent slabs at distances of 0.5, 1.5, 2.5, 3.5, 4.5 and $19.5\, \ell_0$ from the fluctuating membrane surface. Black lines indicate the bulk case. }
\label{fig:std2_ov_t}
\end{figure}

Fig.~\ref{fig:std2_ov_t} displays $D_{\perp,\parallel}(t)$ for different slabs near the fluctuating membrane.
At this level, the features of $D_{\perp,\parallel}(t)$ are similar for the rigid and flat cases (see Appendix~\ref{app:diffusion_flat}) and so we focus on the fluctuating membrane.
In the perpendicular direction, diffusive motion exhibits several stages. The diffusion coefficient $D_{\perp}$ first increases before reaching a quasi plateau around $t\simeq 1-10 \ \tau_0$, corresponding to the onset of the diffusive regime. This time increases slightly with increasing distance from the interface. A molecule starting closer to the interface is more likely to reach a stationary regime sooner as it will more quickly probe the domains closer to the interface. Diffusive motion is then quickly hindered around $t \simeq 10 - 100\ \tau_0$, at a timescale again increasing with increasing distance from the interface. Perpendicular mobility is reduced because perpendicular motion is bounded by the interface, which is again, reached sooner for molecules starting closer to the interface.  The characteristic time scale to diffuse vertically across a slab is $\tau_{\rm slab} = \ell_0^2/D_b^{\infty} \simeq 10 \ \tau_0$, and indeed, for the first slab (darkest color), molecular mobility decays around $\tau_{\rm slab}$. Eventually perpendicular mobility decays to zero at very long times $t\simeq 10^4\  \tau_0 \simeq \tau_{\rm conf}$, due to effective confinement as all molecules sense the interface's periodic image. 

In the parallel direction, $D_{\parallel}$, much like $D_{\perp}$, experiences a quasi-plateau around $t\simeq 1-10\ \tau_0$; this quasi-plateau is more or less clear depending on the distance from the interface. Eventually at long times, as molecules sense the entire vertical domain, $D_{\parallel}$ reaches another plateau which corresponds to $D_{\parallel}^{\infty}$.


\subsection{Spatially-dependent diffusive properties}


To compare the dynamics in different systems, we characterize  Brownian motion by a single diffusion coefficient in each slab. Specifically, we choose to focus on $D_{\perp/\parallel}(t_d)$, taking $\tau_{\rm diff}<t_d< \tau_{\rm slab}$ after the onset of the diffusive regime but before significant diffusion in between slabs. For parallel diffusion we set $t_d = 10\ \tau_0$, and for orthogonal diffusion, $t_d$ corresponds to the maximum of $D_{\perp}(t)$ in each slab, which is $\approx 10\ \tau_0$. 
For simplicity, we write $D_{\perp/\parallel} = D_{\perp/\parallel}(t_d)$. To increase statistical accuracy, the results for $D_{\perp/\parallel}$ are averaged over 10 independent simulations. 

\begin{figure}[th]
\centering
\includegraphics[scale=0.2]{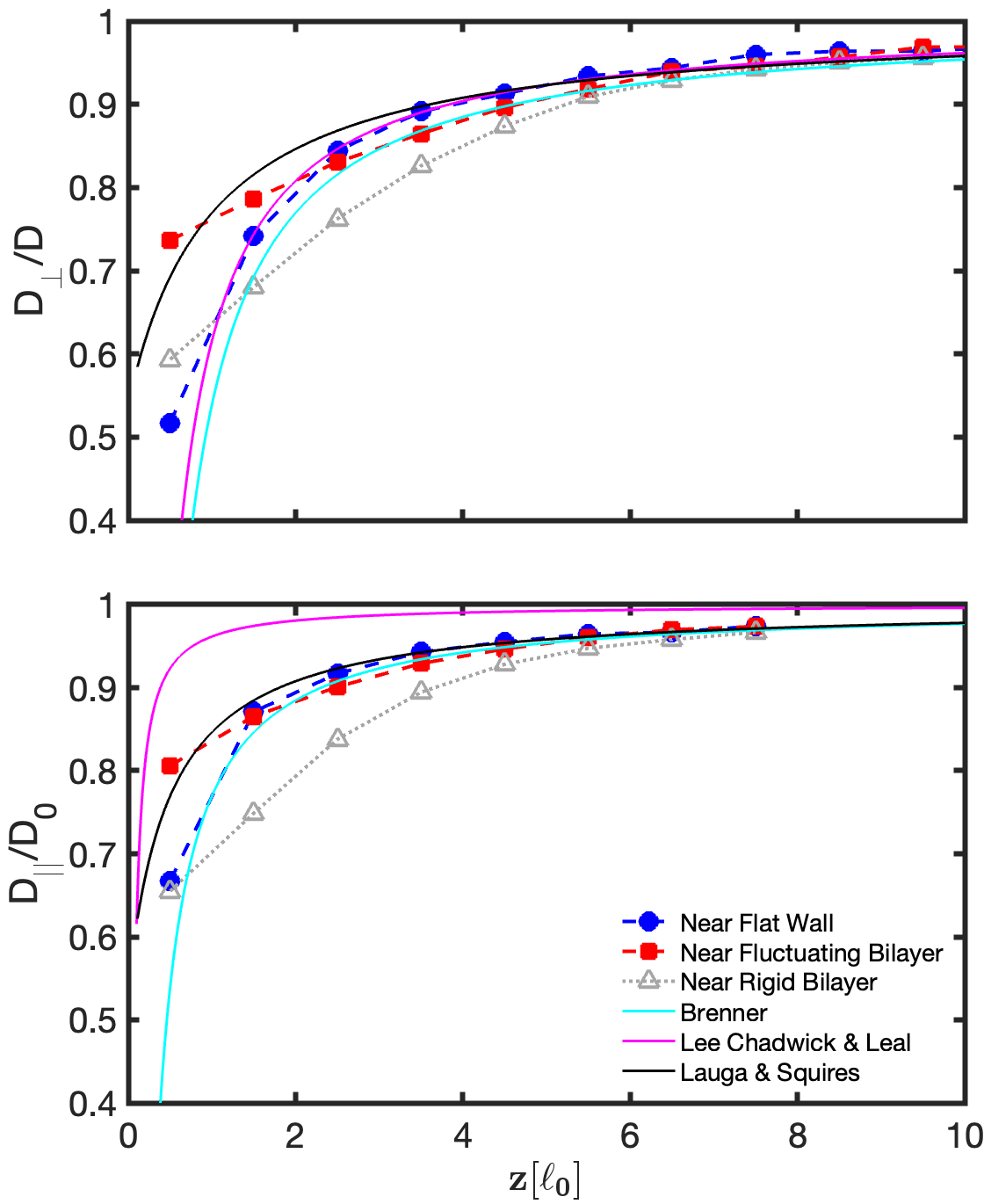}
\caption{Profiles of the diffusion coefficients $D_{\perp}$ and $D_{\parallel}$ as a function of the distance to the interface. Symbols are from the present simulations. Solid cyan (resp. magenta) lines are obtained from Brenner, Eq.~\eqref{eq:lubrication} (resp. Ref.~[\onlinecite{lee1979motion}]) who considered the motion of a solid particle moving near a flat wall (resp. a flat liquid interface). The black lines are obtained from Eq.~\eqref{eq:diff_Lauga} with $\ell_s=\ell_0$. }
\label{fig:diff_profile}
\end{figure}

The diffusion coefficients $D_{\perp/\parallel}$ in each slab at height $z$ above the interface are displayed in Fig.~\ref{fig:diff_profile}.
Close to the interface, both diffusion coefficients are reduced in all systems explored. 
Far from the interface, values approach the bulk diffusion coefficient $\Dbulk$, without being strictly equal to $\Dbulk$ due to the small but finite confinement of the solvent in the $z$ direction. 

A few striking differences should be noted between the different systems. The reduction of $D_{\perp}$ near a flat rigid wall is steep, with a sharp decrease between the 1st and 2nd slabs, and where significant mobility reduction is observed until $\mathcal{L}_{\rm HI} = 4-5 \ \ell_0$. In contrast, the mobility reduction near a rigid or fluctuating membrane is smoother, continuously decreasing over 4-5 slabs, with the same slope including in the 1st and 2nd slabs. $D_{\perp}$ is smallest in the rigid membrane case for nearly the entire range, except in the 1st slab. The mobility in the fluctuating membrane case is always larger than the rigid membrane case, and even larger than near the flat wall on the 1st and 2nd slabs. This is consistent again with a fluidification effect of the fluctuating membrane.
 Similar conclusions can be made in the parallel direction, with however an overall smaller reduction of the diffusion coefficient in that case.

To gain phenomenological insights, the mobility reduction near the flat wall can be compared with familiar, continuum lubrication theories.
Fig.~\ref{fig:diff_profile} displays the diffusion coefficient calculated from Brenner's theory near a hard wall (cyan lines), whose simplified expression reduces to Eq.~\eqref{eq:lubrication} to first order in $a/z$, where $a$ is the molecule's radius. The applicability of Brenner's theory to molecular motion is limited since equations are derived in the continuum limit. Nonetheless, we choose $a = 0.41\ell_0$ as an order of magnitude and compare qualitatively the results. Strikingly, the shape of the prediction, with a sharp decrease near the interface (cyan), is in excellent agreement with both diffusion coefficients $D_{\perp}(z)$ and $D_{\parallel}(z)$ for the flat wall (blue circles). This suggests that qualitative insights on diffusive motion near a flat wall can be transferred from finite size solid particles translating in a viscous fluid to small molecules with no specific physico-chemical interactions. 

The fluidification effect near the fluctuating membrane has similar effects as an apparent slip at the interface. Inspired by the work of Olsen et al.~\cite{olsen2021diffusion} who studied the diffusive motion of lipid nanoparticles near a lipid membrane, we can write the diffusion coefficient in the form of Brenner's law, accounting for an apparent slip length $\ell_s$ at the interface following a theory by Lauga and Squires~\cite{lauga2005brownian} as:
\begin{equation}
\frac{D_\perp(z)}{\Dbulk}=1-\frac{9}{8}\frac{a}{z+\ell_s} \quad \text{and}\quad \frac{D_\parallel(z)}{\Dbulk}=1-\frac{9}{16}\frac{a}{z+\ell_s}.
\label{eq:diff_Lauga}
\end{equation}
The diffusion profiles associated with Eq.~\eqref{eq:diff_Lauga} are plotted in Fig.~\ref{fig:diff_profile} in black, assuming $\ell_s=\ell_0$. The theoretical profiles with slip are significantly flattened near the interface, and agree with simulations of the fluctuating membrane. Molecular motion near a bilayer can thus be described with an effective finite slip. Although the slip length $\ell_s = \ell_0 = 0.64~\mathrm{nm}$ is small compared to measured slip lengths in experimental systems, which are more on the order of $10~\mathrm{nm}$ at least~\cite{lauga2007microfluidics,secchi2016massive}, the slip length scale is relevant in our system as it is similar to that of the characteristic height of membrane corrugations $\delta h=1.3\, \ell_0$. Near the rigid membrane, however, such an approach fails to predict diffusive motion, for any slip length. 

\section{Mechanisms responsible for the fluidification of molecular motion near wiggly interfaces}

We seek to understand the mechanisms underlying our observations in Sec.~\ref{sec:observations}. Overall, the fluidification of motion near the fluctuating membrane could be due to several effects: (A) geometry: the membranes display wells and humps that could locally modify motion; (B) wiggliness: as the membrane moves, it could push groups of solvent molecules, inducing flows that could help mixing~\cite{marbach2018transport,wang2023interactions}; (C) liquid aspect of the membrane: lipids can diffuse freely in the fluctuating case, lowering the stresses at the interface; (D) elasticity of the membrane could induce memory effects and modify friction in various ways~\cite{fares2024observation,daddi2016long}. 
 Our goal is now to test these different mechanisms. 


\subsection{Membrane wells act as traps; the wavy geometry creates an effectively ``smoother interface''.}
\label{sec:displacements}

The density profiles in Figs.~\ref{fig:membrane_density} and~\ref{fig:w_density_profiles} show that molecules near membranes can reside in deeper positions compared to the flat wall. These correspond to membrane wells, located in regions of negative height fluctuation (see Fig.~\ref{fig:mb_snapshot}). Several of our observations suggest geometry, \textit{i.e.} the presence of wells and bumps in a membrane, hinders motion: indeed, in the rigid membrane case, both $D_\parallel^{\infty}$ and the local mobility $D_{\parallel,\perp}(z)$ are reduced compared to the flat and fluctuating cases. We hypothesize that this is due to a ``trapping effect''. Many theories demonstrated that the long-time diffusion coefficient of particles should be decreased by the constrictions which form effective ``entropic'' barriers, where a particle spends a long time exploring wells before it can escape that well~\cite{zwanzig1992diffusion,jacobs_diffusion_1967,reguera_kinetic_2001, kalinay_corrections_2006,rubi2019entropic}. This effect has only been proven to act on $D_\parallel^{\infty}$ but here we find it, naturally, also has an effect on local mobility $D_{\perp, \parallel}(z)$. 

\begin{figure}[th]
\centering
\includegraphics[scale=0.22]{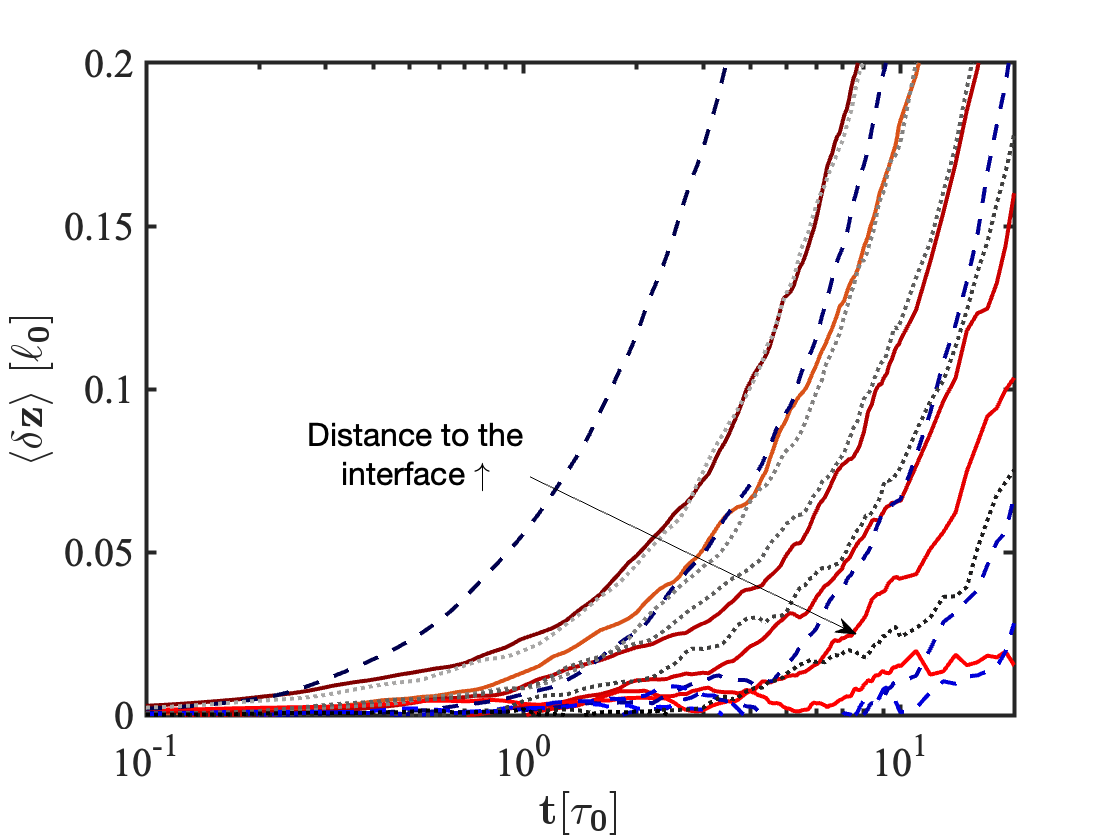}
\caption{Mean perpendicular displacement of solvent molecules  with time. Same color code as Fig.~\ref{fig:std2_ov_t}.}
\label{fig:meanDelta_t}
\end{figure}

To add further proof to this trapping hypothesis, we analyze orthogonal displacements relative to the membrane, $\langle \delta z(t) \rangle = \langle z(t) - z(0)\rangle$, for each slab, see  Fig.~\ref{fig:meanDelta_t}. 
The mean displacement remains positive in time, indicating a clear repulsion of molecules away from the interface. The closer the solvent molecules to the interface, the earlier and the stronger the repulsion. Near the flat wall (blue), the mean displacement is larger in the first two layers than near the membranes (gray/red). Similar conclusions may be drawn from the probability distribution functions of vertical displacements (see Appendix~\ref{app:vertical}).
The displacement being comparable near both membranes is consistent with the presence of wells close to the interface where molecules stay for long times and are not repelled as efficiently far from the interface. This increases local friction, and explains the discrepancies in the mobility measurements between the rigid membrane and the flat wall cases. 
It also illustrates that measurements of dynamic properties near a membrane are averaged over the lateral direction, and so averaged over wells and bumps, creating an effectively ``smoother'' landscape vertically. 

Geometry thus has 2 effects: (i) membrane wells act as traps -- molecular drift away from the interface is slowed down compared to near a flat surface -- increasing local and global friction and (ii) the wavy geometry creates an effectively ``smoother interface''.
\subsection{Dynamical fluctuations kick solvent molecules}
\label{sec:mb_fluct}

Although solvent molecules near the fluctuating membrane should experience increased friction due to membrane geometry, like near the rigid membrane, motion is clearly facilitated in this fluctuating case. Compared to other cases, the fluctuating membrane continuously experiences fluctuations of thermal origin that emerge and relax. 
This is consistent with a transfer of momentum hypothesis, where the motion of the membrane itself agitates molecules and facilitates mixing. 

\begin{figure}[h!]
\centering
\includegraphics[scale=0.6]{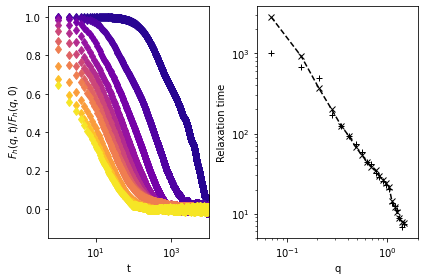}
\put(-250,0){(a)}
\put(-110,0){(b)}
\caption{a) Time decay of the autocorrelation function $F(q,t)$, for modes with wavenumber $q = n_q q_1$ where $q_1=0.069\ell_0^{-1}$ is the smallest wavenumber and $n_q$ denotes odd numbers between 1 and 21 (from purple to yellow). 
b) Relaxation time $\tau_r(q)$ of the membrane fluctuations, calculated from Eq.~\eqref{eq:taur}. $+$ and $\times$ symbols correspond to simulations lasting for $10^4\ \tau_0$ and $15\ \times 10^4\tau_0$, respectively, demonstrating the profile is converged. 
Times are given in $\tau_0$ units and $q$ as $1/\ell_0$. 
} 
\label{fig:mb_Fqt_relax_time}
\end{figure}

To test this hypothesis, we first characterize the relaxation dynamics of the fluctuating membrane and then see how they relate to molecular diffusion.
For each fluctuating mode $q$ in Fourier space, we calculate the height autocorrelation function $F(q,t) = \langle \tilde{h}(q,t) \tilde{h}^\star(q,0) \rangle$ where $\tilde{h}(q,t)$ is the Fourier transform of $h(x,t)$ (see Appendix~\ref{app:mb_dynamics}). The decay of $F(q,t)$ for different modes $q$ shown in Fig.~\ref{fig:mb_Fqt_relax_time}a is not purely exponential. This trend has been observed in previous lipid bilayer studies, and described via stretched exponential~\cite{brandt2010stretched} or double exponential~\cite{seifert1993viscous,shkulipa2006simulations} models. To simplify, we characterize $F(q,t)$ by a ``global'' relaxation time scale~\cite{brandt2010stretched}, as 
\begin{equation}
    \tau_r(q)=2 \displaystyle \int_0^\infty\frac{F(q,t)^2}{F(q,0)^2} dt.
    \label{eq:taur}
\end{equation} 
The relaxation time $\tau_r(q)$ decays with increasing wavenumber $q$, see Fig.~\ref{fig:mb_Fqt_relax_time}b, reaching values between $10 -10^3 \ \tau_0$.

How does this relaxation time relate compare with other relevant timescales?
As the membrane fluctuations occur over multiple length scales, inspired by Marbach et al.~\cite{marbach2018transport}, we build a dimensionless P\'eclet number as the ratio of a characteristic diffusive time scale $\tau_{\rm diff}(q) = 1/\Dbulk q^2$ across a characteristic length scale $1/q$ compared with the relaxation time at that scale
$$\mathrm{Pe} = \frac{\tau_{\rm diff}(q)}{\tau_r(q)} = \frac{1}{\Dbulk q^2 \tau_r(q)}.$$ 
We use $\Dbulk \simeq 0.1 \ \ell_0^2/\tau_0$ as an order of magnitude, and reading off Fig.~\ref{fig:mb_Fqt_relax_time}b, we find membrane fluctuations typically exhibit $\mathrm{Pe} \simeq 1$, independently of the fluctuation length scale. 
Dynamic membrane fluctuations thus act on similar time and length scales as purely diffusive motion and thus can potentially affect overall mobility. 

Convective mixing induced by dynamic fluctuations has been shown theoretically~\cite{marbach2018transport} to modify long time parallel diffusion in confined media $D_{\parallel}^{\infty}$. When the P\'eclet number $\mathrm{Pe} \gtrsim 1$ (respectively $\mathrm{Pe} \lesssim 1$), $D_{\parallel}^{\infty}$ is enhanced (respectively decreased) compared to a flat wall case. For $\mathrm{Pe}\simeq 1$, the outcome is not obvious, and thus it is remarkable that fluctuations here are sufficient to enhance $D_{\parallel}^{\infty}$ compared to the flat wall case. The rigid membrane case corresponds to infinite relaxation times $\tau(q) = \infty$, thus to $Pe\rightarrow 0$, and reduced diffusion, which is consistent with our study, and with the ``trapping effect'' of geometry. 
In this theory, the factor $2 \delta h^2/H^2$ quantifies the change in diffusion coefficient where $\delta h$ is the amplitude of the interface corrugations and $H$ the distance between the confining interfaces. Here $\delta h = 1.3\ \ell_0$ and $H = 46\ \ell_0$ giving a relative change of 0.04\%, which is much smaller than the observed differences of about 2\% in Table~\ref{table:Db_confined}. The discrepancies between the theory and our observations are likely due to molecular effects, since Ref.~[\onlinecite{marbach2018transport}] is only valid for much larger particles than the solvent. Yet, qualitatively the effects are similar: purely spatial corrugations of the confining domain reduce motion by trapping them, while local membrane motion kicks molecules and facilitates their motion. 

The theory in  Ref.~[\onlinecite{marbach2018transport}] does not provide insights on local diffusion, so as to explain the trends in Fig.~\ref{fig:diff_profile}. Nonetheless, we can still expect similar effects are at play and well described by the P\'eclet number.
For large wavenumbers, $\tau_r(q\simeq 1 \, \ell_0^{-1}) \simeq 10\ \tau_0$, which is of the order of the diffusion time scale of the solvent molecules over a slab $\tau_{\rm slab}$. Thus, a molecule diffusing near the membrane, experiences several local wiggly movements of the membrane before moving away. This coupling between the molecular motion and the quick membrane fluctuations likely explains the fluidification effect of diffusive motion at small $z$ near the fluctuating interface.


\subsection{Membrane fluidity itself is likely not at play}

Another major difference between the rigid and fluctuating membrane is that lipids are free to diffuse in the fluctuating  case. This could possibly facilitate lateral motion by reducing local stress. The lipid molecules are much larger than the solvent molecules, and the ratio of the lipid lateral diffusion coefficient within the membrane ($D_l$) and the solvent diffusion coefficient ($\Dbulk$) is quite small, $D_l/\Dbulk \simeq 0.01$, in agreement with previous findings~\cite{GrootRabone2001}. This suggest that lipid fluidity operates on timescales that are much larger than that of local diffusion, and hence are likely playing a minor role. 

To push this reasoning slightly further, we present in Fig.~\ref{fig:diff_profile} a theory (in magenta) corresponding to the motion of a sphere near a non-deformable liquid-liquid interface, from Lee et al.~\cite{lee1979motion}, where
\begin{equation}
    \frac{D_\perp(z)}{\Dbulk} \simeq 1- \frac{15}{16} \frac{a}{z} \quad \text{and} \quad \frac{D_\parallel(z)}{\Dbulk} \simeq 1- \frac{3}{32} \frac{a}{z}.
    \label{eq:lee}
\end{equation}
The predictions using Eq.~\eqref{eq:lee} yield similar profiles for $D_{\perp}(z)$ versus Brenner's with the same radius, but quite different for $D_{\parallel}(z)$, demonstrating lateral fluidity should primarily facilitate only lateral motion. In contrast, the behavior for all simulated cases are quite similar in the tangential and perpendicular directions. The fluidification effects we observe in the fluctuating case are thus different from stress reduction mechanisms near a fluid interface.

\subsection{Lack of elastic effects in the diffusive regime}
\label{sec:force}

Finally, we investigate whether the molecules near the membrane experience elastic forces originating from the deformability of the fluctuating membrane.
Inspired by Ref.~[\onlinecite{fares2024observation}], we assume that at distance $z$ from the interface, the mean forces on a molecule can be decomposed into a thermal force associated with the equilibrium density distribution $F_{\rm eq}(z)$ -- reflecting intermolecular interactions --, a drag force resisting molecule drift away from the interface $F_{\rm drag}(z)$, spurious drift arising from the coupling between random motion and the dependence of molecule mobility on the distance to the interface $F_{\rm mn}(z)$, and finally a non-conservative interaction force $F_{\rm int}(z)$ with the interface. The different contributions originate from an overdamped Langevin equation for particle motion. Our simulations allow us to measure all \textit{conservative} forces, \textit{i.e.} all forces but $F_{\rm int}(z)$. With these measurements we demonstrate $F_{\rm eq}(z) + F_{\rm drag}(z) + F_{\rm mn}(z) = 0$ (see Fig.~\ref{fig:drift_forces} and details in Appendix~\ref{app:force_balance}) and so the absence of non-conservative forces in our case. 

Soft non-conservative forces have been observed experimentally for a colloidal drop near a wall~\cite{fares2024observation} and theoretically for a particle near an elastic membrane~\cite{daddi2016long}. However, we do not observe them here. This apparent paradox relates to the size of the diffusing object and the time scale of observation.  Indeed, such soft forces are only present at ``short'' timescales. In Ref.~[\onlinecite{fares2024observation}] the non conservative forces appeared at a time scale of the order of $0.01-1~\mathrm{s}$ for a micron-sized particle. 
The diffusion coefficient near the interface is $D_{\parallel} \simeq 0.1~\mathrm{\mu m^2/s}$ which amounts to a diffusion timescale across the particle size of about 
$3~\mathrm{s}$. Therefore, in the case of molecular diffusion near the fluctuating membrane, if present, the non-conservative forces are expected to occur at timescales shorter than the time to diffuse across a molecule's diameter, and so in the ballistic regime. They thus have no effect in the diffusive regime. 

\section{Discussion and concluding remarks}
\begin{figure}[th]
\centering
\includegraphics[scale=0.45]{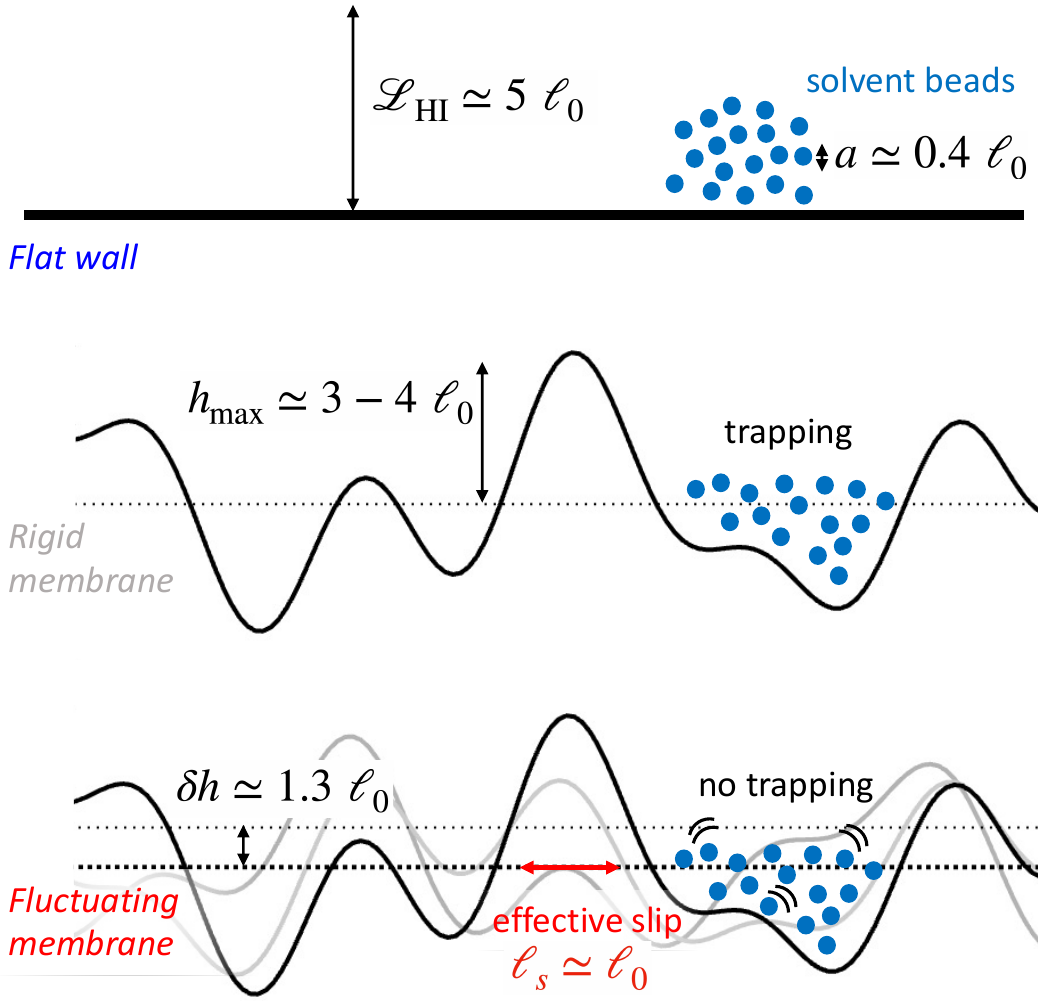}
\caption{Schematic summary of the mechanisms influencing molecular mobility near the different interfaces, and associated length scales. Solvent molecules are trapped in the wells of the rigid membrane, while dynamic membrane fluctuations help them explore beyond the wells. See text for details. }
\label{fig:graph_abst}
\end{figure}

We have probed numerically the diffusive motion of solvent molecules, in the parallel and perpendicular directions near 3 different interfaces. Our observations are best recapitulated by noticing the different length scales at play, which correspond to the different mechanisms unveiled -- see Fig.~\ref{fig:graph_abst}.

Near a flat wall, as reference, we found that the molecules local diffusion coefficient is reduced, and that this reduction can be described by continuum theories for the mobility reduction of a solid particle of size $a$ near a flat solid wall. Here we estimated the characteristic radius $a$ of a DPD molecule -- typically representing 3 ``real'' water molecules -- as half the distance at which the radial pair distribution function of the solvent peaks.  The ``hydrodynamic'' length scale measuring the distance over which mobility reduction is observed is $\mathcal{L}_{\rm HI}\approx 5\ \ell_0\approx 12\ a$.

In comparison, we explored  molecular motion near a membrane made of lipid self-assembly. The membrane average height fluctuations with respect to the membrane midplane are $\delta h=1.3\ \ell_0$, whereas local maxima or minima of the membrane reach $h_{\rm max}\approx 3-4\ \ell_0$. 
Near the rigid, or frozen in time, membrane, we found a reduction of local solvent mobility and of long-time parallel diffusion, compared to the rigid case. 
We suggest that the membrane's wavy geometry, creates wells that act as local traps for motion, significantly reducing mobility compared to a flat case. Since $h_{\rm max} \simeq \mathcal{L}_{\rm HI}$ we can expect that the hydrodynamic mobility reduction occurs over a similar length scale as the ``trapping'' effect, which was indeed measured in our case. Overall, trapping reduces the long-time parallel diffusion as well.

In the fluctuating case, when the membrane experiences spontaneous thermal motion, we found local mobility is comparable to the flat wall case, and even facilitated in the layer that is the closest to the membrane. We estimated the length scale of this fluidification effect through an effective slip length $\ell_s\approx\ell_0$, which is close to the characteristic magnitude of height fluctuations $\delta h$. Through an evaluation of the ratio between the time to diffuse across a wave, and the relaxation time of that wave, we found that dynamic kicks of the fluctuating membrane increase molecular agitation. This conclusion is conformed by the remarkable similarity between $\ell_s$ and $\delta h$, showing that fluidification is likely promoted over a length scale that is comparable to the mean membrane fluctuations. 
Several phenomena are not expected to be at play, such as lipid mobility within the membrane or elastic effects which respectively occur on timescales that are much longer or shorter than the timescales relevant for diffusive motion at this molecular level.




While we have probed individual dynamic properties, such as the self-diffusion coefficient of the molecules, it is natural to ask how other dynamic properties may be modified near soft fluctuating interfaces.
For instance, a recent numerical investigation demonstrated that the flexibility of a porous matrix can allow for enhanced permeability by deforming the local structure~\cite{schlaich2024bridging}. Other ``fluctuating aspects'' such as longitudinal vibration modes of a flat surface (phonons) have been shown theoretically and experimentally to modify hydrodynamic friction and electronic currents in the surface~\cite{coquinot2023quantum,lizee2023strong}. Soft and/or fluctuating interfaces thus exhibit exotic consequences on varied dynamical properties.

\begin{figure}[th]
\centering
\includegraphics[scale=0.22]{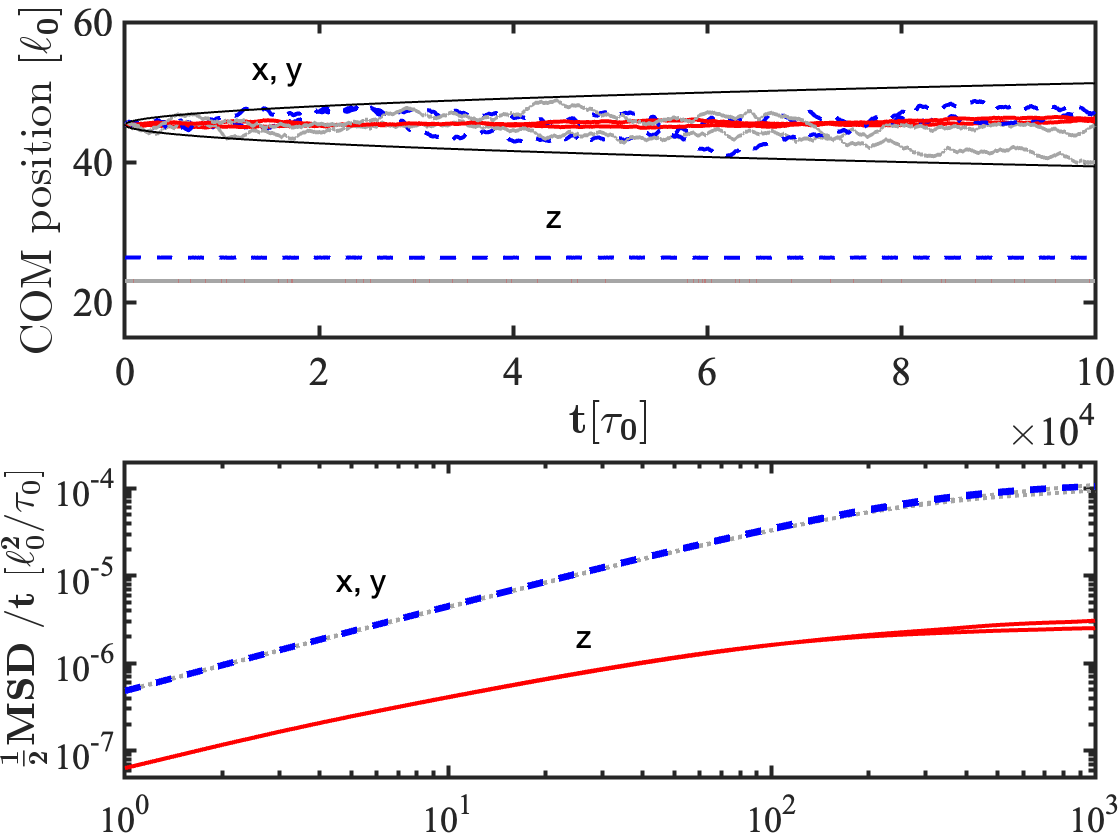}
\put(-215,170){(a)}
\put(-215,72){(b)}
\caption {Temporal evolution of a) the solvent center of mass and b) its half MSD over time, near a fluctuating membrane (solid red), a rigid membrane (dotted gray) and a flat wall (dashed blue). The direction of motion is indicated on the figure. The black lines in a) correspond to $y = y_0 +\sqrt{2D_{\rm DB}t}$. }
\label{fig:COM_long_water}
\end{figure}

As an illustration, we briefly investigate in our system a canonical collective dynamic quantity, the center of mass (COM) of the solvent, see Fig.~\ref{fig:COM_long_water}. In the parallel direction, the COM diffuses in the domain bounded by the flat wall or the rigid membrane, whereas it is significantly slowed down in the fluctuating membrane case. This is surprising as we have found individual diffusion is mostly accelerated by the presence of fluctuations. The center of mass diffusion coefficient in a slit can be modeled~\cite{detcheverry2013thermal} as
$D_{\rm com}= k_B T H /12 \eta L W  = 1.6 \times 10^{-4} \ \ell_0^2/\tau_0$ and
Fig.~\ref{fig:COM_long_water} displays, in black, the theoretical position of the COM as $y = y_0 +\sqrt{2D_{\rm com}t}$. This prediction is in remarkable agreement with the flat and rigid membrane cases, for which we can estimate $D_{\rm com} \simeq  10^{-4} \ \ell_0^2/\tau_0$. The drastic slow down induced by fluctuations could be hypothesized to be due to the effective slip. Adding slip boundary conditions on the interface modifies the COM diffusion coefficient~\cite{detcheverry2013thermal}, by a factor
$\left( 1+6\ell_s/H\right)$. 
This factor is small, since $\ell_s \simeq \ell_0$, and it is always greater than 1, thus contributing to enhance diffusion, opposite to the observed trend. Another phenomenon must be at play! Collective dynamic properties near fluctuating interfaces thus hold many more mysteries to be rationalized.


This work brings some insights into apparent models of the motion of fluids confined by a soft interface, where the diffusion, and subsequently the viscosity, would depend on the distance to the interface. A still open question of interest is 
the impact of interface fluctuations on the friction coefficient or slip of the fluid at the interface~\cite{bocquet2013green}, and how this couples with the change in diffusion.

\begin{acknowledgments}
The authors acknowledge fruitful discussions
with Benoit Coasne, Amael Obliger, Jean-Louis Barrat and Aleksandar Donev.
This work was granted access to the HPC resources of IDRIS under the allocations 2024-[A0152A12031] made by GENCI, and to computational resources provided by the computing meso-centre CALMIP under project No.~P1002. S.M. and M.A. acknowledge financial support from the CNRS through the MITI interdisciplinary program (MembranesJumelles). A.M and Z.L. acknowledge the financial support from the National Science Foundation via grants OAC-2103967 and CDS\&E-2204011.
\end{acknowledgments}

\appendix

\section{Lipid and water simulation model}\label{app:coarse_grain}

\begin{figure}[h!]
    \centering
    \includegraphics[scale=0.23]{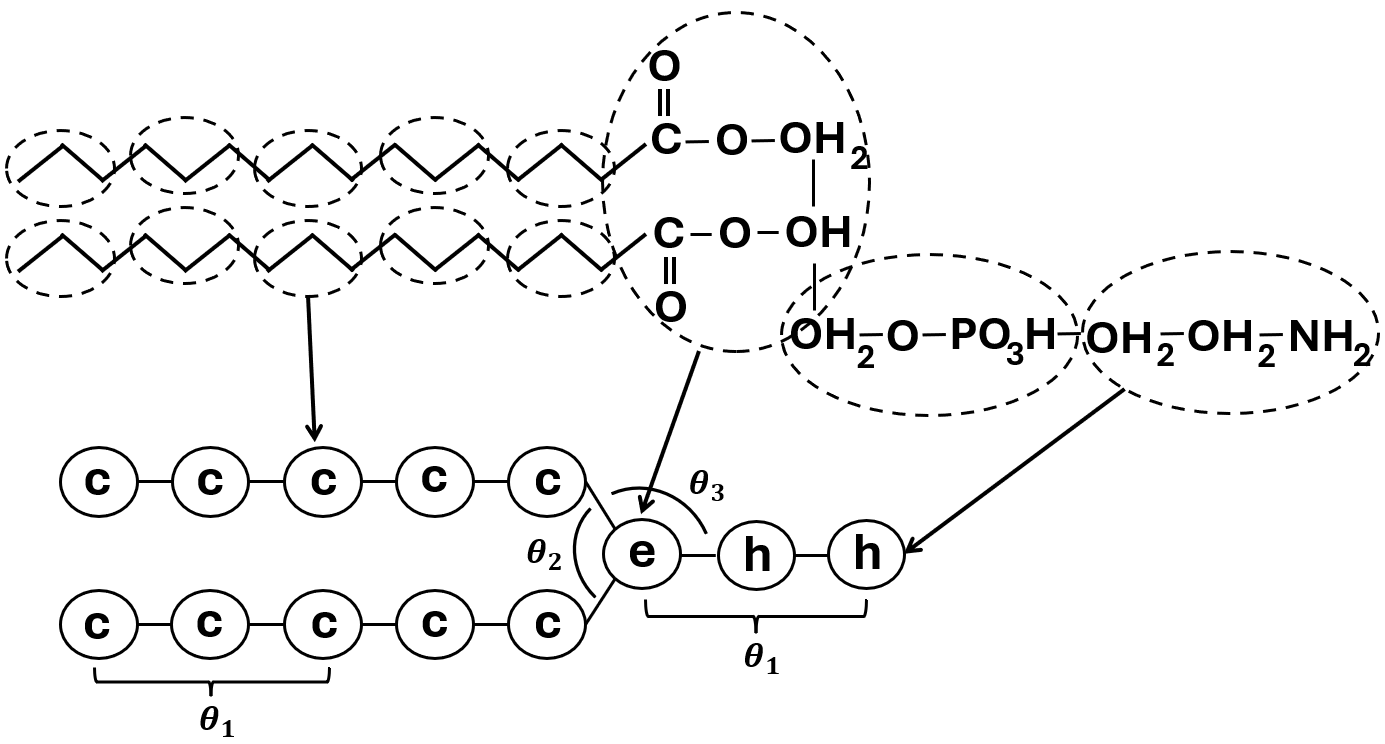}
    \caption{Phosphatidylethanolamine (lipid) molecule and its coarse-grained representation.}
    \label{fig:dpd_mapping}
\end{figure}

The system we model comprises phosphatidylethanolamine (PE) and water. 
The structure of PE consists of two \( \text{C}_{15} \) hydrocarbon chains attached to a glycerol group.
Following the approach of Groot and Rabone~\cite{GrootRabone2001}, 3 carbon atoms are grouped into one DPD bead of type \text{\textbf{c}}, while each bead of type \text{\textbf{e}} represents 1.5 glycerol-linking unit. The PE molecule and its mapping to the coarse-grained DPD model are shown in Fig.~\ref{fig:dpd_mapping}. The association between the numerical and physical systems is mainly done through the number density of water molecules which is given by \(\rho = N_A \rho_w / M_w\), where \(N_A\) is the Avogadro constant, and \(\rho_w\) and \(M_w\) are the density and molar mass of water, respectively.

Based on the partial volumes of \((\text{CH}_2)_3\) and the glycerol-linking molecule, as reported by Lu et al.~\cite{lu1993neutron}, Groot and Rabone determined that both type \text{\textbf{c}} and type \text{\textbf{e}} DPD beads correspond to a volume of \(90 \, \text{\AA}^3\). The volume of a water molecule is approximately \(30 \, \text{\AA}^3\); thus, each water DPD bead (type \(\text{\textbf{w}}\)) represents three water molecules. 

With the number density of DPD beads \(\rho = 3\), and the coarse-graining parameter \(N_m = 3\), the reference length scale of the system \( \ell_0 = \sqrt[3]{N_m \times \rho \times 30} \, \text{\AA} = 6.46 \text{\AA}\). The mass of each DPD bead is given by \( m= N_m \times \frac{M_w}{N_A} \approx 9 \times 10^{-26} \) kg, where \( M_w = 18 \) g/mol is the molar mass of water, and \( N_A = 6.022 \times 10^{23} \) mol\(^{-1}\) is Avogadro number. Given the reference temperature \(T = 300 \) K, and Boltzmann constant \({k_B} = 1.38 \times 10^{-23}\) J/K, the reference time scale \(\tau_0 = \ell_0 \sqrt{m/{k_B} T} = 3 \) ps.


\section{Dissipative Particle Dynamics method}\label{app:dpd}

In Dissipative Particle Dynamics, each DPD bead \(i\) is characterized by its position and momentum, where time evolution is governed by momentum and energy conservation, as described by the following set of equations~\cite{groot1997dissipative}:
\begin{align*}
\frac{d\mathbf{r}_i}{dt} &= \mathbf{v}_i, \\
m_i\frac{d\mathbf{v}_i}{dt} = \mathbf{F}_i &= \sum_{j \neq i} \left( \mathbf{F}^C_{ij} + \mathbf{F}^D_{ij} + \mathbf{F}^R_{ij} \right).
\end{align*}
Here, \(m_i\) represents the mass of bead \(i\), \(t\) represents time, while \(\textbf{r}_i\), \(\textbf{v}_i\), and \(\textbf{F}_i\) denote the position, velocity, and force vectors of bead \(i\), respectively. The summation is performed on all other beads within a cutoff radius $r_c \equiv \ell_0$. The three pairwise components of ${\textbf{F}_i}$, the conservative force $\mathbf{F}^C_{ij}$, the dissipative force $\mathbf{F}^D_{ij}$, and the random force $\mathbf{F}^R_{ij}$ are given by\cite{groot1997dissipative}:
\begin{align*}\displaystyle\mathbf{F}_{ij}^C &= a_{ij}\omega_C(r_{ij})\textbf{e}_{ij}, \\
\displaystyle \textbf{F}_{ij}^D &= -\gamma_{ij}\omega_D(r_{ij})(\textbf{e}_{ij} \cdot \textbf{v}_{ij})\textbf{e}_{ij}, \\
\displaystyle \textbf{F}_{ij}^R &= \sigma_{ij}\omega_R(r_{ij})\xi_{ij}\Delta t^{-1/2}\textbf{e}_{ij}. 
\end{align*}
Here, \(r_{ij}\) = |\({\textbf{r}_{ij}}\)| is the distance between the beads \(i\) and \(j\), \({\textbf{e}_{ij}}\) the unit vector in the direction of \({\textbf{r}_{ij}}\), \(\textbf{v}_{ij}\) is the relative velocity of the bead \(i\) with respect to the bead \(j\) and \(\Delta t\) is the time step. \(a_{ij}\) is a repulsive force parameter. For similar beads, \(a_{ii}\) can be determined by matching the compressibility of the fluid using the following~\cite{pivkin2010dissipative}:
\begin{equation}
a_{{ii}} = k_B T (\kappa_c^{-1}N_m - 1.0)/2 \alpha \rho,
\label{eq:a_ij}
\end{equation}
where \(\alpha = 0.101 \pm 0.001\) for liquid water which arises from the DPD equation of state~\cite{groot1997dissipative}. \(\kappa_c^{-1}\) is the dimensionless compressibility of the fluid, given by~\cite{groot1997dissipative} 
\begin{equation}
\kappa_c^{-1} = \frac{1}{\rho k_B T \kappa_T},
\label{eq:dimensionless_kappa}
\end{equation}
 where $\rho$ is the number density of the molecules and \(\kappa_T\) is the isothermal
 compressibility of the fluid.  For beads of different types, the values of \(a_{ij}\) are determined to reproduce the mutual solubility of different species by matching the Flory–Huggins \(\chi\)-parameters~\cite{groot1997dissipative}. 
\(\xi_{ij}(t)\) is a randomly fluctuating variable with Gaussian statistics, such that \(\langle \xi_{ij}(t) \rangle = 0\), \(\text{Var}(\xi_{ij}) = 1\), and \(\langle \xi_{ij}(t) \xi_{kl}(t') \rangle = (\delta_{ik}\delta_{jl}+\delta_{il}\delta_{jk})\delta(t-t')\).
 The dissipative force and the random force jointly act as a thermostat provided that the dissipative parameter \(\gamma_{ij}\) and the amplitude of the noise \(\sigma_{ij}\) satisfy the fluctuation-dissipation equilibrium, which requires \(\sigma^2_{ij} = 4 \gamma_{ij} k_B T\) and \(\omega_D(r_{ij}) = \omega_R^2(r_{ij})\). A common choice for the weight functions is \(\omega_C(r_{ij}) = 1 - r_{ij} / r_c\) and \(\omega_D(r_{ij}) = (1 - r_{ij} / r_c)^s\) for \( r \leq r_C \) and zero for \( r > r_c \), where $r_c$ is the cutoff radius. 
The exponent \(s\) can be modified to adjust the kinematic viscosity and diffusivity of the liquid, thereby achieving a desired Schmidt number~\cite{li2014energy}. It is common to use $s=2$ following the original DPD formulation, but in this work, we consider $s=0.5$ which allows to reach $Sc\approx 10$ (instead of $Sc\approx 1$ for $s=2$)~\cite{li2014energy}.

Additional bond interactions are necessary to link lipid chain beads. 
A commonly used approach is applying translational (\( U_T \)) and angular (\( U_\theta \)) harmonic spring
potentials, as follows:
\begin{align*}
U_T &= \frac{1}{2} k_T (r - r_{eq})^2,\\
U_\theta &= \frac{1}{2} k_\theta (\theta - \theta_{eq})^2.
\end{align*}
Here, \(r_{eq}\) and \(\theta_{eq}\) denote the equilibrium bond length and angle, respectively, while \(k_T\) and \(k_\theta\) represent the spring constant and angular bending stiffness.

\section{DPD Parameters}\label{app:dpd_param}

The isothermal compressibility of water at atmospheric pressure and reference temperature \(T = 300 \, \text{K}\) is \(\kappa_T = 4.5 \times 10^{-10} \, \text{Pa}^{-1}\) (see Ref.~[\onlinecite{kell1970isothermal}]). Using Eq.~\eqref{eq:dimensionless_kappa}, we calculate \(\kappa_c^{-1} = 16.01[k_BT/\ell_0^3]\). Consequently, from Eq.~\eqref{eq:a_ij}, we determine the interaction coefficient \(a_{ww} = 78\) for the water beads. The \(a_{ij},\) parameters for the other species are adopted from Groot and Rabon~\cite{groot2001mesoscopic} and summarized in Table~\ref{table:a_xy}. 

\begin{table}[h!]
    \centering
    \renewcommand{\arraystretch}{1.0} 
    \setlength{\tabcolsep}{13pt} 
    \caption{Repulsive $a_{ij}$ interaction parameter between the species.}
    \begin{tabular}{c c c c c}
        \hline\hline
        bead type & w & c & e & t \\ \hline
        w & 78  & 104 & 79.3 & 79.3 \\
        c & 104 & 78 & 86.7 & 104 \\
        e & 79.3 & 86.7 & 78 & 79.3 \\
        h & 75.8 & 104 & 79.3 & 86.7 \\
        \hline\hline
    \end{tabular}
    \label{table:a_xy}
\end{table}

The dissipative parameter \(\gamma_{ij} = 4.5\) and the exponent \(s = 0.5\) are used for all bead interactions. The cut-off radius \(r_c = 1\) is used for all species and interactions.

For the bonded beads, the linear spring constant is \( k_T = 4.0 \), and the equilibrium bond length is \( r_{eq} = 0 \). The bending stiffness is \( k_\theta = 6.0 \), with equilibrium bond angles \(\theta_1 = 180^\circ\), \(\theta_2 = 90^\circ\), and \(\theta_3 = 135^\circ\).


\section{Bulk Fluid Properties}\label{app:bulk_prop}

Here, we characterize the dynamic and structural properties of the solvent molecules (\text{\textbf{z}} type beads) in the bulk. 

\begin{figure}[h!]
    \centering
    \includegraphics[scale=0.325]{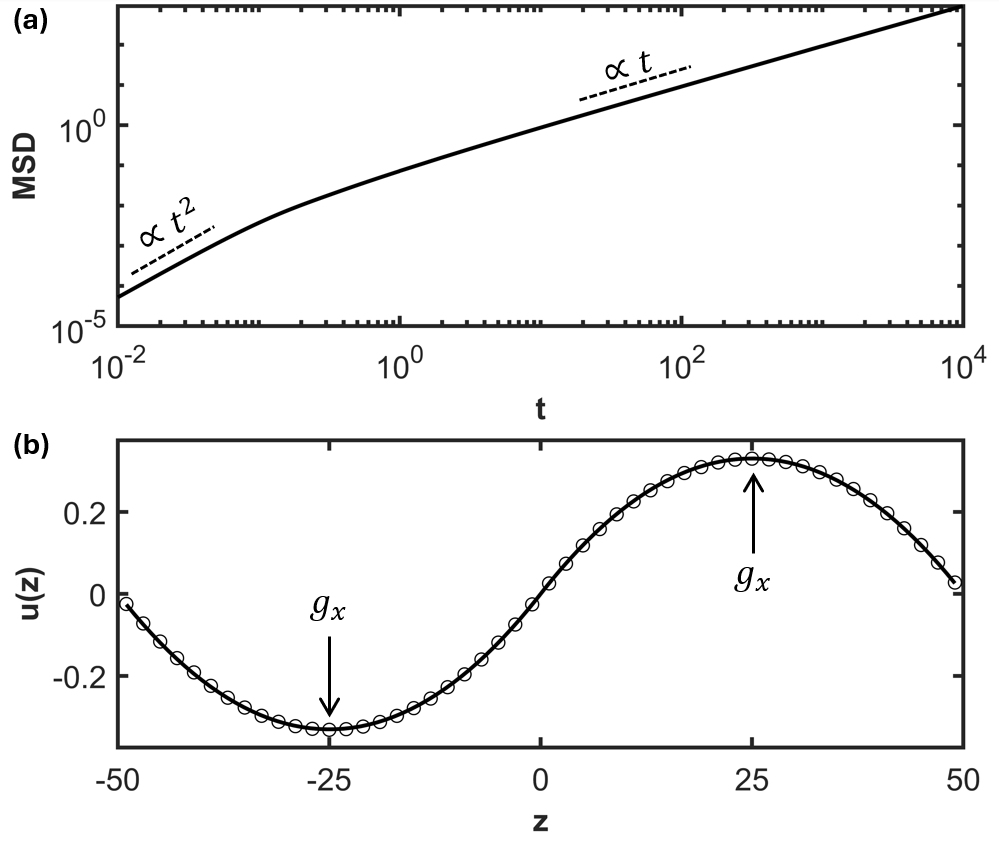}
    \caption{(a) Time evolution of the mean-squared displacement (divided by 6) of the solvent molecules. The time and MSD are indicated in $\tau_0$ and $\ell_0^2$ unit, respectively.
    (b) Velocity profile obtained from the Poiseuille flow simulation (circles). The coordinate $z$ and the velocity are indicated in $\ell_0$ and $\ell_0/\tau_O$ units, respectively. The line is a fit using  Eq.~\eqref{eq:dpflow}. The computational domain here is \((L_x\times L_y\times L_z)=(10\times 100\times 100)\ell_0^3\). }
    \label{fig:dpflow_msd0}
\end{figure}

\paragraph{Self-diffusion coefficient.} The self-diffusivity of the fluid \(\Dbulk\) is determined in the absence of an interface, from the mean-squared displacement of the DPD beads, 
which is displayed in Fig.~\ref{fig:dpflow_msd0}(a) as a function of time.
After an initial ballistic regime, motion is diffusive (\(\text{MSD} \propto t\)) for $t \gtrsim 0.1 \tau_0$. The slope of the line in the diffusive regime represents the self-diffusivity of the fluid, which we can quantify to be $D_b = 0.092 \ \ell_0^2/\tau_0$.

\paragraph{Viscosity.}  The kinematic viscosity is determined using a periodic Poiseuille flow method~\cite{backer2005poiseuille} by fitting the analytical solution 
\begin{equation}
u(z) = g_x z(d - |z|)/2\nu
\label{eq:dpflow}
\end{equation} 
to the velocity profile obtained from sheared DPD simulations. Here, \(\nu\) is the kinematic viscosity, \(g_x = 0.001 \ell_0/\tau_0^2\) is the body force per unit mass applied to each DPD bead, and \(d = 50 \ell_0\) is half the length of the computational domain in the \(z\)-direction. The velocity profile obtained from the DPD simulation and the fitted analytical solution are shown in Fig.~\ref{fig:dpflow_msd0}(b).
 We find \(\nu = 0.94\ \ell_0^2/\tau_0\). 

\begin{figure}[h!]
    \centering
    \includegraphics[scale=0.55]{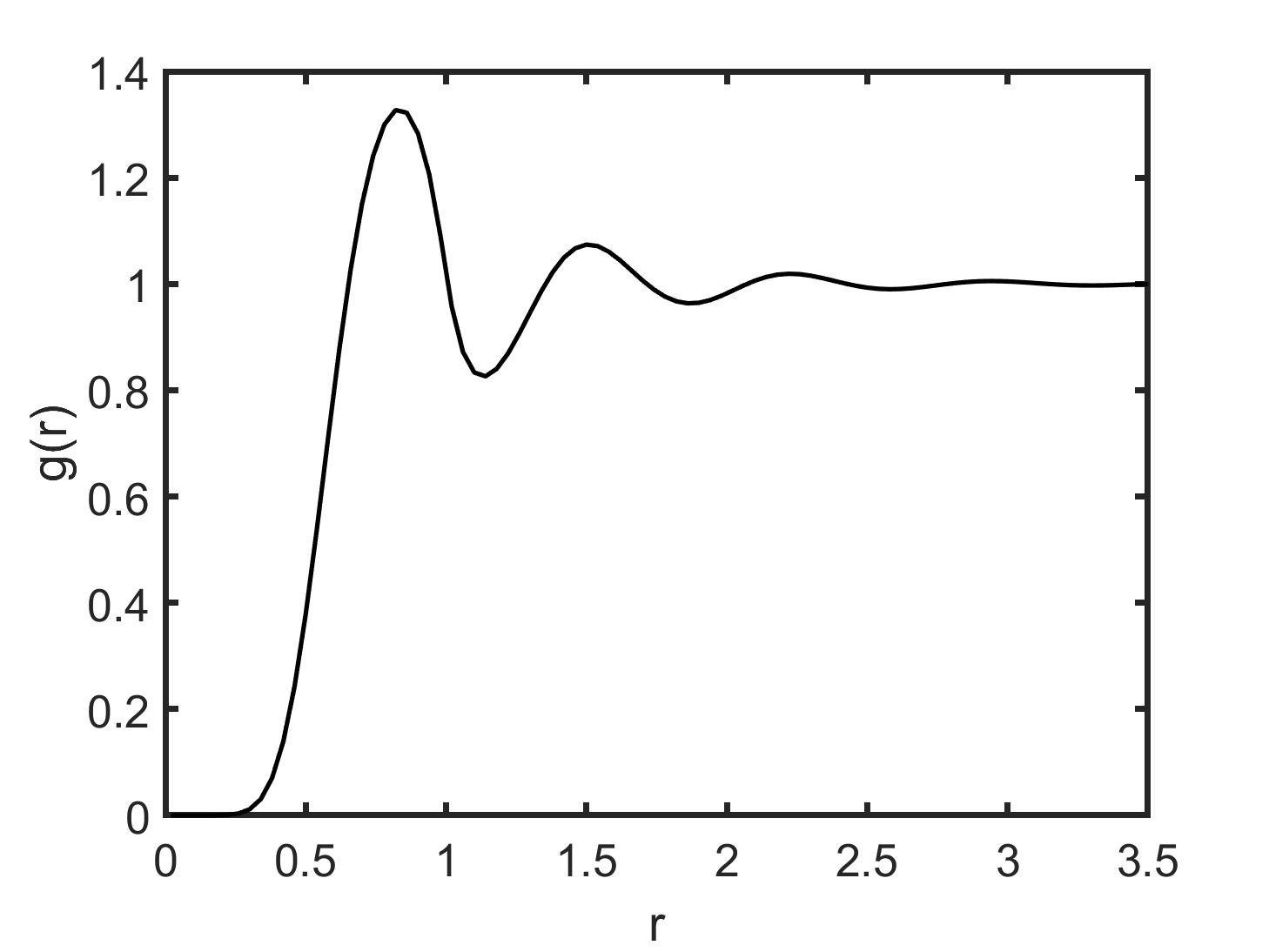}
    \caption{Radial pair distribution function of the solvent molecules. The distance $r$ is indicated in $\ell_0$ unit.}
    \label{fig:rdf}
\end{figure}

\paragraph{Pair distribution function.}  Another quantity of interest is the characteristic distance between the DPD molecules. The unique length scale prescribed in the DPD method is the cut-off distance of the pairwise forces, which here we take to be the same for all force contributions: conservative, dissipative and random. This length scale is not necessarily representative of the distance between molecules. Instead, we find the characteristic distance between the DPD molecules (or points) from the distance at which the largest peak of the radial pair distribution function occurs, here $0.82\ \ell_0$ -- see Fig.~\ref{fig:rdf}. Thus, we determine a characteristic bead "radius" equal to half this distance, \textit{i.e.} $a=0.41\ \ell_0$. 

\section{Diffusive solvent dynamics near flat and undulated rigid membranes}\label{app:diffusion_flat}

\begin{figure}[h!]
\centering
\includegraphics[scale=0.22]{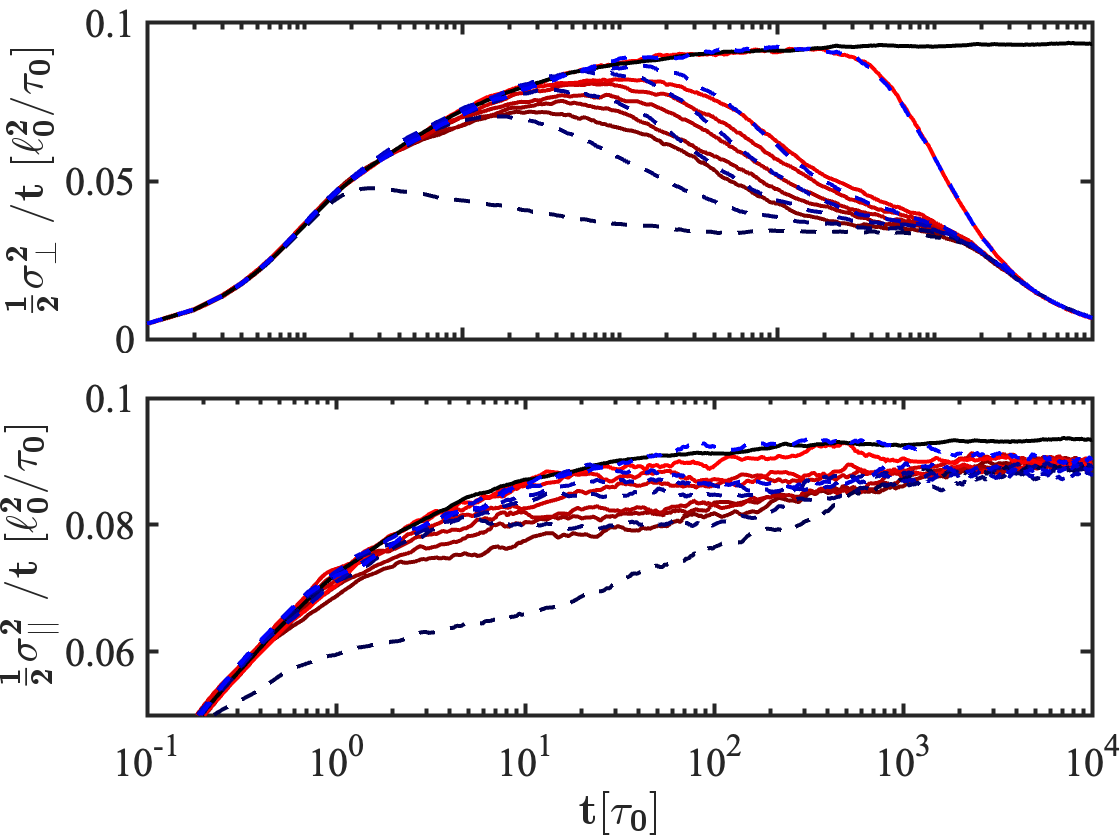}
\put(-210,170){(a)}
\put(-210,85){(b)}
\caption{Comparison of the time evolution of $\frac{1}{2}  \frac{\sigma_{\perp,\parallel}^2(t)}{t}$ in the (a) perpendicular and (b) parallel directions, between simulations with a fluctuating membrane (red) and a rigid wall (dashed blue). Red or blue lines, from darkest to lightest, represent slabs at distances of 0.5, 1.5, 2.5, 3.5, 4.5 and $19.5\, \ell_0$ from the fluctuating membrane surface. Black lines indicate the bulk case.}
\label{fig:std_effect_of_shape}
\end{figure}

\begin{figure}[h!]
\centering
\includegraphics[scale=0.22]{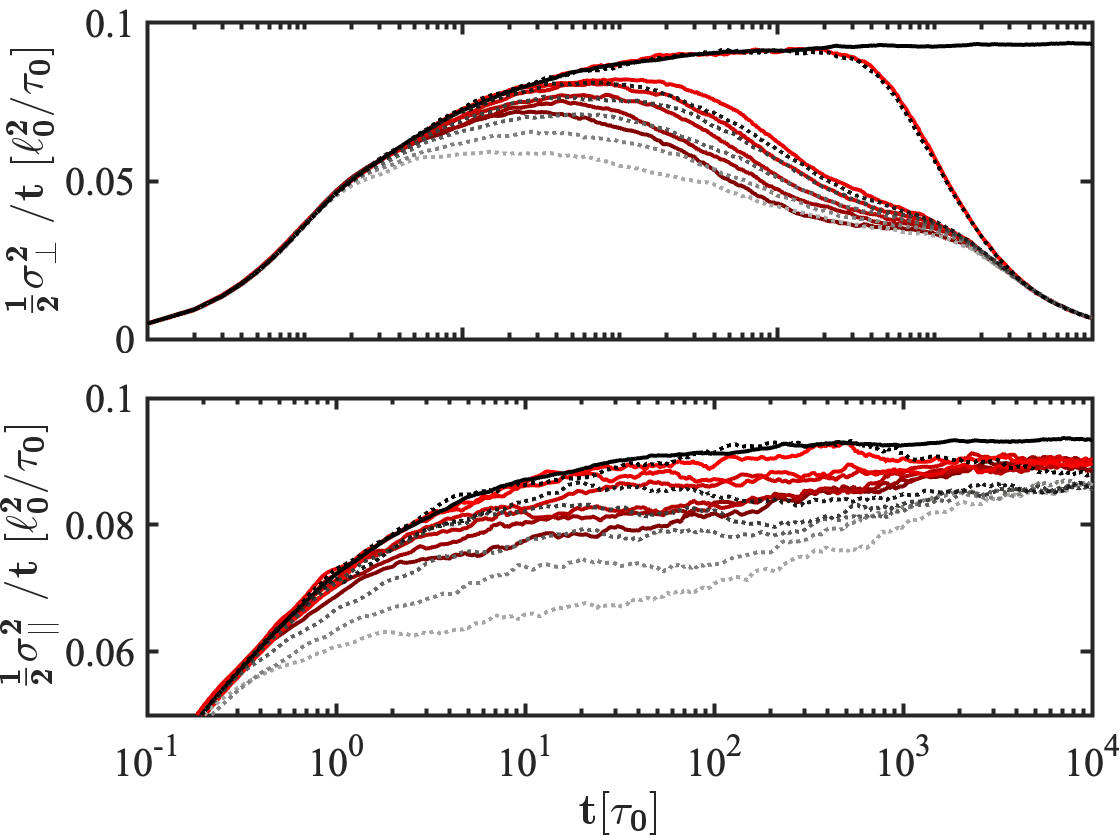}
\put(-210,170){(a)}
\put(-210,87){(b)}
\caption{Comparison of the time evolution of $\frac{1}{2}  \frac{\sigma_{\perp,\parallel}^2(t)}{t}$ in the (a) perpendicular and (b) parallel directions, between simulations with a fluctuating membrane (red) and a rigid membrane (dotted gray). Red or blue lines, from darkest to lightest, represent slabs at distances of 0.5, 1.5, 2.5, 3.5, 4.5 and $19.5\, \ell_0$ from the fluctuating membrane surface. Black lines indicate the bulk case.}
\label{fig:std_effect_of_rigidity}
\end{figure}

In Figs.~\ref{fig:std_effect_of_shape} and~\ref{fig:std_effect_of_rigidity} we report the standard deviation of solvent motion in the different slabs near the flat and undulated rigid membranes respectively, with the fluctuating membrane case superimposed for comparison. 

\section{Vertical displacements of solvent molecules}\label{app:vertical}

\begin{figure}[h!]
\centering
\includegraphics[scale=0.22]{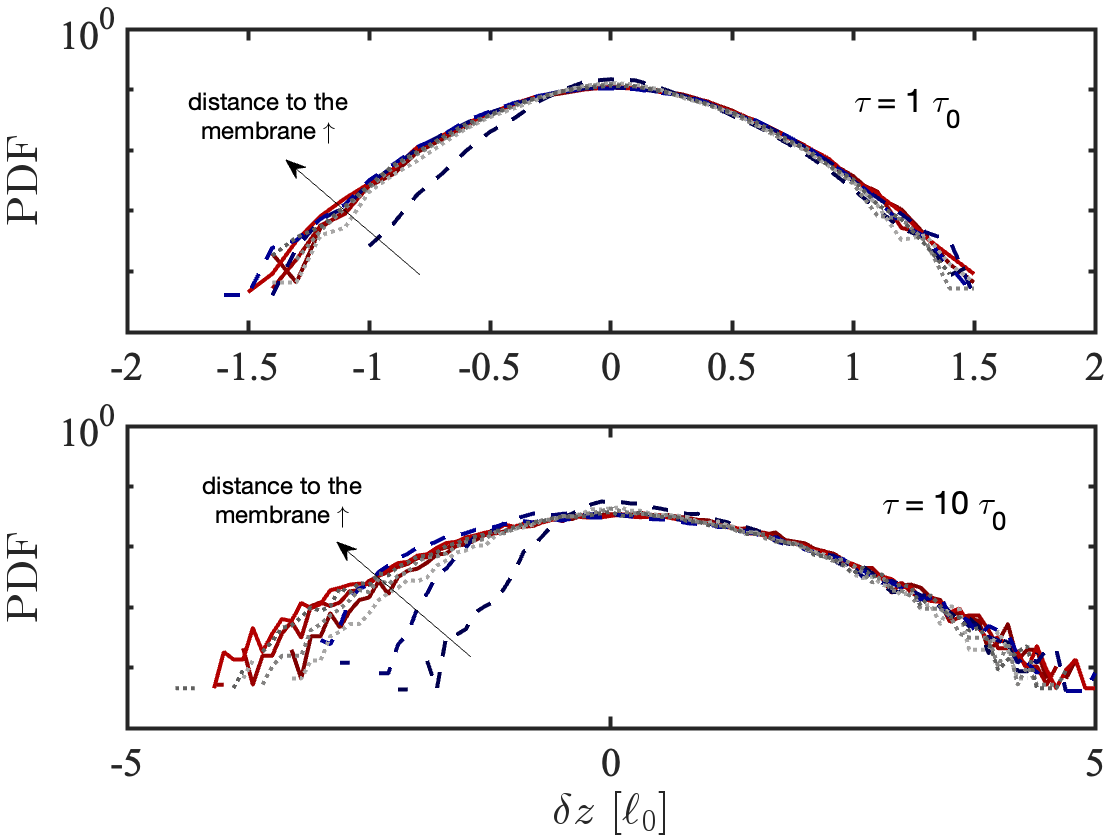}
\put(-215,170){(a)}
\put(-215,82){(b)}
\caption{Normalized probability density function $P(\delta z)$ of vertical bead displacements $\delta z$ at different time points (a) $t=\tau_0$ and (b) $t=10\ \tau_0$, at different distances from the different interfaces: fluctuating membrane (red), flat wall (blue) and rigid membrane (gray). Dark to light colors represent slabs at distances 0.5, 1.5, and $2.5\ \ell_0$ with respect to the membrane surface.}
\label{fig:w_pdf_dx}
\end{figure}

We investigate here the probability distribution functions (PDF) of molecular displacement $P(\delta z)$. Fig.~\ref{fig:w_pdf_dx}a shows the PDF of vertical displacements at time $t= \tau_0$ which roughly corresponds to the end of the ballistic regime. At this early time, the PDF curves remain symmetric (Gaussian) at different distances from the different interfaces, except very close to the flat wall. There, we notice a slight asymmetry in the PDF, towards positive (upward) displacements. This is due to the repulsion of molecules by the rigid flat wall. 

Fig.~\ref{fig:w_pdf_dx}b shows the PDF of vertical displacements at time $t=10\ \tau_0$ which corresponds roughly to the molecular diffusion time scale. We recall that $t = 10\ \tau_0$ also corresponds to the time it takes for the solvent molecules to diffuse along a distance equal to the slab height $\ell_0$. The asymmetry of the PDF between negative displacements (toward the interface) and positive displacements (away from the interface) is apparent for different layers and on different interfaces. Clearly, in the rigid flat wall case, repulsion is felt quite strongly on 3 layers above the wall. In contrast, in both rigid and fluctuating membrane cases, there is only a mild repulsion by the walls, and mostly on the 2 first layers. This corroborates the hypothesis that the wavy nature of the interface traps molecules near the interface. 

Note that the PDF of the longitudinal displacements (not shown here) remained symmetric and did not show significant time dependence, regardless of the nature of the interface and the distance to it. 

\section{Membrane dynamics}\label{app:mb_dynamics}

The intensity of membrane height fluctuations, as well as the rate of their relaxation in time depend on the wavenumber. Several post-processing steps are required to characterize the membrane relaxation time scales. First, the lipid beads in the membrane are distributed on a $32\times 32$ two-dimensional grid (parallel to the membrane). Second, the membrane height $h(\bm{x},t)$ on each grid point $\bm{x}$ is given by the average of the $z$ coordinates of the beads in that grid point. $z=0$ is set at the time-averaged position of the membrane midplane. Third, the power spectrum $\langle \left|h_q^2(t)\right| \rangle$ of the height fluctuation of each undulating mode with wavenumber $q$ is calculated. Here $\langle \cdot \rangle$ is an average over time only. Fourth, the time autocorrelation function of the height fluctuations is computed,
$F(q,t)=\left<\left|h_q\right|(t)\left|h_q\right|(0)\right>$. The decay in time of the autocorrelation function for different $q$ is shown in Fig.~\ref{fig:mb_Fqt_relax_time}a.

When integrating $F(q,t)$ to calculate the relaxation time $\tau_r(q)$ from Eq.~\eqref{eq:taur}, we set a cutoff value on $t$, such that $F(q,t)/F(q,0) > 0.05$ to avoid integrating noise. A smaller cutoff of 0.1 has no significant influence on the calculated relaxation time. 

\section{Force balance on the solvent slabs}\label{app:force_balance}

A force balance in the $z$ direction (normal to the membrane), originating from an overdamped Langevin equation for particle motion, can be written as follows:  
\begin{equation}
    F_{\rm eq}(z) + F_{\rm drag}(z) + F_{\rm mn}(z) + F_{\rm int}(z)  = 0,
    \label{eq:forcebalance}
\end{equation}
where the term 
$$\displaystyle F_{\rm eq}(z)=\frac{\mathrm{d}U_{\rm eq}}{\mathrm{d}z}$$
results from equilibrium interactions with the membrane through an interaction potential $U_{\rm eq}(z)$. It can be inferred from the density profile of the solvent molecules at equilibrium, which obeys Boltzmann distribution and so $U_{\rm eq}(z)=-k_BT \ln(\rho(z)/\rho)$. 

The second term $F_{\rm drag}(z)$ corresponds to the Stokes drag experienced by the solvent molecules while they drift away from the interface. The drag force can be modeled using the drift velocity of the solvent molecules and considering that the bead mobility is reduced near the interface 
$$\displaystyle F_{\rm drag}(z)=-\gamma(z) v_{\perp} \equiv -\frac{k_BT}{D_\perp(z)}v_{\perp},$$ 
where $\gamma(z) = k_B T/D_{\perp}(z)$ is the friction coefficient, and $v_{\perp}$ is the velocity of the molecules perpendicular to the interface. This velocity was estimated from a time average between $t=0$ and $t=10\ \tau_0$ of the instantaneous velocity $\frac{<\delta z(t+\delta t)> - <\delta z(t)>}{\delta t}$ with $\delta t=0.1 \ \tau_0$.

The third term $F_{\rm mn}(z)$ corresponds to multiplicative noise or spurious drift 
$$F_{\rm mn}(z)=\displaystyle \frac{k_BT}{D_\perp(z)} \frac{\mathrm{d}D_\perp}{\mathrm{d}z}$$
and arises from the coupling between the random motion and the dependence of molecule mobility on the distance to the interface. Like the classic drag, this contribution is a resisting force. We can understand it as a resistance against the collective diffusion of the molecules from regions of larger diffusion toward regions of lower diffusion. The derivative of the diffusion coefficient was calculated as $\frac{\mathrm{d}D_\perp}{\mathrm{d}z}=\frac{D_{\perp,i+1}-D_{\perp,i}}{z_{i+1}-z_i}$, where $i$ refers to the slab index and $z_i$ is the position of the center of the slab.

The last term $F_{\rm int}(z)$ denotes a force that may result from the interaction between the freely moving particles and the interface, either due to electrostatic forces if present or due to elastic effects associated with membrane fluctuations in the present study.

\begin{figure}[th]
\centering
\includegraphics[scale=0.22]{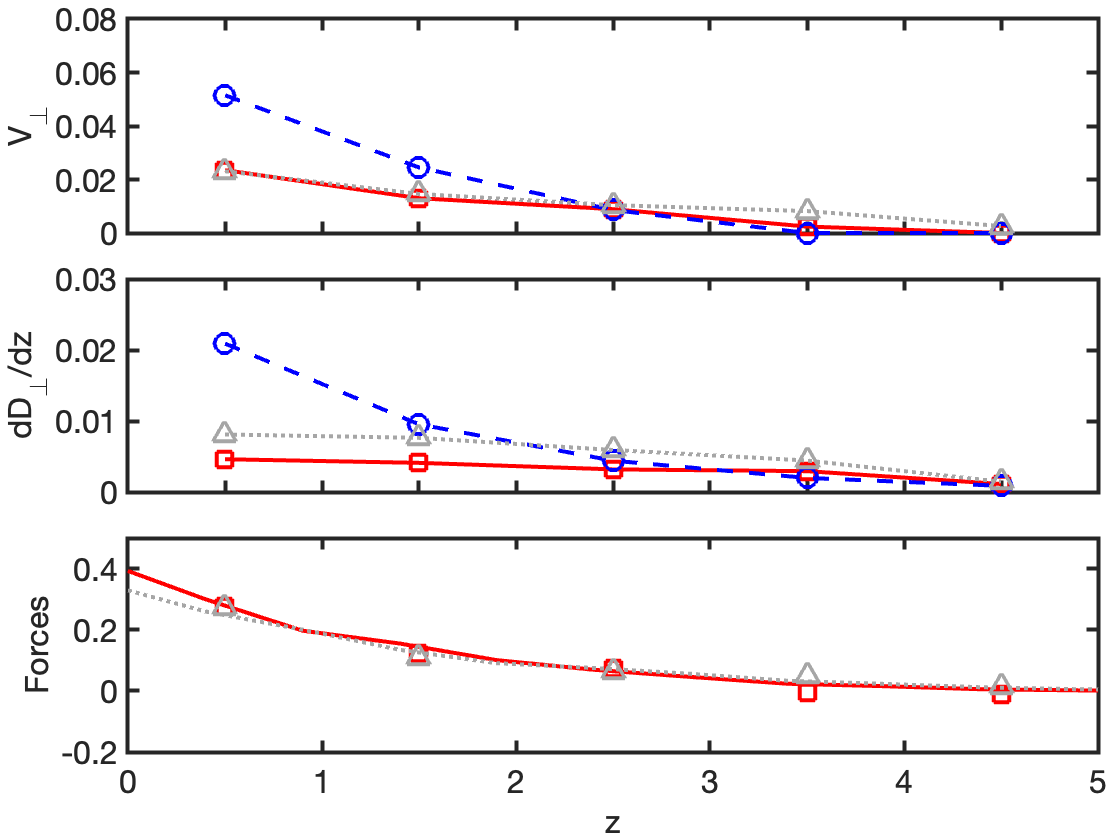}
\put(-20,170){(a)}
\put(-20,110){(b)}
\put(-20,55){(c)}
\caption{a) Average drift velocity, b) derivative of the normal diffusion coefficient and c) forces experienced by the solvent molecules as a function of the distance to the interface. In a) and b), red squares, blue circles and gray triangles are obtained from simulations with a fluctuating membrane, a flat wall and a rigid membrane, respectively, while the lines are used for eye guidance.  In c), red squares and gray triangles correspond to 
$-(F_{\rm drag}+F_{\rm mn})$ whereas red solid and gray dotted lines correspond to $F_{\rm eq}$,
obtained from simulations with a fluctuating (red) and a rigid membrane (gray). The coordinate $z$ and the forces are indicated in $\ell_0$ and $\sqrt{k_BT}/\ell_0$ units respectively, whereas the derivative of the diffusion coefficient and the drift velocity are indicated in $\ell_0/\tau_0$ units. }
\label{fig:drift_forces}
\end{figure}

Our simulations allow us to measure numerically three contributions to the force experienced by the solvent molecules, $F_{\rm eq}(z)$, $F_{\rm drag}(z)$ and $F_{\rm mn}(z)$ and compare them. 
Fig.~\ref{fig:drift_forces}a shows, in blue, $v_{\perp}$, the drift velocity of the solvent molecules as a function of the distance to the interface. The drift is stronger near the flat wall compared to the membrane cases, in agreement with the observations made on particle displacements in Sec.~\ref{sec:displacements}. The decay of $v_\perp$ occurs at short length scales, typically around $3-4\ \ell_0$, which is comparable to the maximum height fluctuations (displayed in Fig.~\ref{fig:mb_snapshot}).
Fig.~\ref{fig:drift_forces}b displays the spurious drift velocity associated with the gradient of the molecules' mobility $\mathrm{d}D_\perp/\mathrm{d}z$.
The amplitude of this spurious drift is smaller than the amplitude of the average drift for the three interfaces tested here. 
This is consistent with the presence of a repulsive interaction potential which should also contribute a force in the positive vertical direction. The spurious drift also decays, in a similar manner as the drift, perpendicularly to the interface. Strikingly, both membranes exhibit only a very weak contribution to spurious drift, due to the nearly flat diffusive profile $D_\perp(z)$ (see Fig.~\ref{fig:diff_profile}a), in agreement with the ``smoothing'' effect associated with averaging along each membrane's surface over wells and bumps. 

Now we can examine the different terms in the force balance. $F_{\rm drag}$ and $F_{\rm mn}$ are of the same nature, \textit{i.e.} they are both written in the form of a product of a molecule's mobility and a relative velocity. However, they apply in opposite directions: the first points towards the membrane unlike the second which points away from it. We lump them in a single term which corresponds to the effective drag. Their negative sum $-(F_{\rm drag}+F_{\rm mn}) = k_BT/D_{\perp}(z) (v_\perp - \mathrm{d}D_\perp(z)/\mathrm{d}z)$ is shown in Fig.~\ref{fig:drift_forces}c, using symbols. 
The continuous lines correspond to $F_{\rm eq}$, using the density profiles displayed in Fig.~\ref{fig:w_density_profiles}. Fig.~\ref{fig:drift_forces}c clearly shows that the effective drag and the equilibrium force are well balanced, for both the fluctuating and rigid membrane cases. Note that similar conclusions can be made in the case of a flat wall (not shown here), with the difference that $F_{\rm eq}$ is not a smooth function in that case, due to the sharp decay of the equilibrium bead distribution. 
We conclude that non-conservative or elastic forces are absent, at least during the diffusive time range.

\providecommand{\noopsort}[1]{}\providecommand{\singleletter}[1]{#1}%

\end{document}